\documentclass{iopart}

\usepackage{graphicx}
\usepackage{dcolumn}
\usepackage{bm}
\usepackage{epstopdf}
\begin{document}

\title{Electronic structure of the substitutional vacancy in graphene:
Density-functional and Green's function studies}

\author{B R K Nanda*, M Sherafati, Z S Popovi\'c** and S Satpathy}%
\address{
Department of Physics $\&$ Astronomy, University of Missouri,
Columbia, MO 65211}%
\ead{satpathys@missouri.edu}

\date{\today}

\begin{abstract}

We study the electronic structure of graphene with a single substitutional vacancy using a combination
of the density-functional, tight-binding, and impurity Green's function approaches.
Density functional studies are performed with the all-electron spin-polarized linear
augmented plane wave (LAPW)  method.
The three $sp^2 \sigma$ dangling bonds adjacent to the vacancy introduce localized states (V$\sigma$) in the mid-gap region, which split due to the crystal field and a Jahn-Teller distortion,  while the $p_z \pi$ states introduce a sharp resonance state  (V$\pi$) in the band structure.
For a planar structure, symmetry strictly forbids hybridization between the $\sigma$ and the $\pi$ states, so that these bands are clearly identifiable in the calculated band structure.
As for the magnetic moment of the vacancy, the Hund's-rule coupling aligns the spins of the four localized V$\sigma_1 \uparrow \downarrow$, V$\sigma_2 \uparrow $, and the V$\pi \uparrow$ electrons resulting in a $S=1$ state, with a magnetic moment of $2 \mu_B$, which is reduced  by about $0.3 \mu_B$ due to the anti-ferromagnetic spin-polarization of the $\pi$ band itinerant states in the vicinity of the vacancy. This results in the net magnetic moment of $1.7 \mu_B$.
Using the  Lippmann-Schwinger equation, we reproduce the  well-known $\sim 1/r$ decay of the localized V$\pi$ wave function with distance and in addition find an interference term  coming from  the two Dirac points, previously unnoticed in the literature.  The long-range nature of the V$\pi$ wave function is a unique feature of the graphene vacancy and we suggest that this may be one of the reasons for the widely varying relaxed structures and  magnetic moments reported from the supercell band calculations in the literature.

\end{abstract}
\submitto{\NJP}
\maketitle

\section{Introduction}

Graphene is a material of considerable interest on account of its unusual linearly-dispersive Dirac band structure and particle-hole symmetry.\cite{Review1, Review2}   Vacancy constitutes  an important defect  center, the  electronic structure of which forms the basic foundation  for the understanding of the behavior of more complex defects including impurities.
Recently it has been suggested that transition-metal doped graphene with vacancies may have potential application in hydrogen storage.\cite{Das}
Experimentally, vacancies in graphene have been created intentionally by irradiating materials with electrons and ions\cite{Hashimoto,  Briddon1, Matilla, Butz} and they may also occur in small concentration during the growth process.\cite{Kushmeric}
While an ideal graphene sheet is non-magnetic, experimental observation of magnetism in carbon systems has been long explained in terms of a variety of defects including isolated vacancies, vacancy clusters, or presence of internal or external boundaries as in nanoribbons.\cite{Butz, Garcia, Ugeda,Cervenka}

\begin{figure}
\centering
\includegraphics[width=5.0cm]{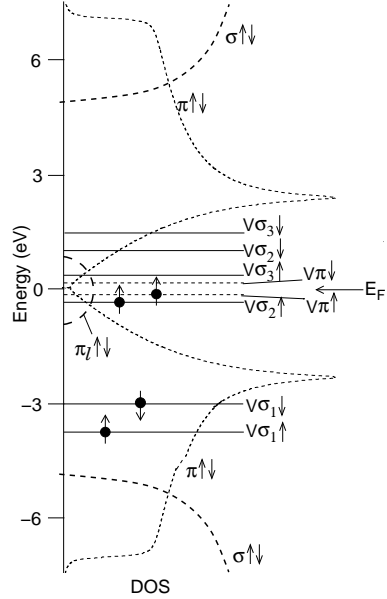}
\caption{Sketch of the electronic structure for an isolated substitutional vacancy in graphene.
The continuum $\pi$ and $\sigma$ bands are shown as  dashed curves, while the vacancy-induced localized  states, V$\sigma$ and V$\pi$, are denoted by straight lines. $E_F$ is the Fermi energy.  The occupied vacancy states are indicated by solid circles with a corresponding net magnetic moment of $2 \mu_B$.   The circular density-of-states (DOS) in the midgap region, labelled $\pi_l \uparrow \downarrow$, indicates schematically the anti-ferromagnetic spin-polarization of the
$\pi$ electron states in the {\it local} neighborhood of the vacancy.  This spin polarization  is responsible for the reduction of the localized magnetic moment from $2 \mu_B$  ($S = 1$) to about $1.7 \mu_B$ in our density-functional calculation.
}
\label{sketchdos}
\end{figure}

There have been several theoretical studies of the isolated vacancy  in graphene from first-principles density-functional theory (DFT)\cite{Briddon,Yazyev, Choi, Singh, Yang, Faccio, Heggie,  nieminen,  ma,   Lim, Dai} or Hartree-Fock calculations\cite{Forte} as well as from tight-binding models\cite{castroneto, hjort, Brey}.  There is also an enormous amount of related work on the chemisorbed defects such as the hydrogen defects and chemisorbed magnetic atoms.\cite{NJP}
Most of the tight-binding models have focused on the $\pi$ bands only, which is clearly inadequate due to the formation of the $sp^2\sigma$ dangling bonds in the mid-gap region. The first-principles calculations do include all relevant states in the band structure including the $sp^2\sigma$ states; however, in spite of all these works, a clear picture of the vacancy states has not emerged.


In this paper, we study the electronic structure of the  vacancy in graphene using the all-electron density functional linear augmented plane waves (LAPW) method along with tight-binding studies as well as the impurity Green's function (GF) approach to interpret the band structure. To our knowledge, this is the first all-electron density functional calculation for the vacancy in graphene reported in the literature.
We have already reported the electronic structure for the mono and bilayer graphenes using the same method.\cite{brk-graphene}
In addition to the DFT calculations, the nature of the vacancy-induced states is modeled from the tight-binding and Green's function studies, which help interpret the DFT results.

\begin{figure}
\centering
\includegraphics[width=8cm]{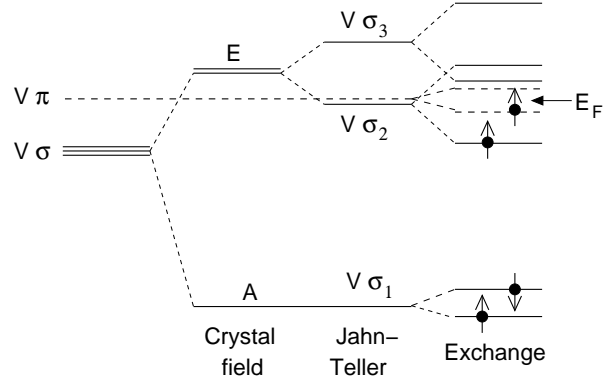}
\caption{Splitting of the three dangling bond $sp^2\sigma$ states of the carbon triangle, denoted by V$\sigma$, and the vacancy-induced zero-mode V$\pi$ state originating from the $\pi$ band. The splitting of the V$\sigma$ states is discussed in detail in Section \ref{sec:Vsigma}.
}
\label{bandsplit}
\end{figure}

The basic overall picture of the electronic structure that emerges from our work is summarized in Fig. \ref{sketchdos}. It shows the standard $\sigma$ and $\pi$ bands of graphene plus the vacancy-induced states, denoted by V$\pi$ and V$\sigma$, which are split due to the crystal field, Jahn-Teller, and the Hund's-rule interactions. The V$\sigma$ states are made out of the three $sp^2\sigma$ dangling bond states, which are located on the three carbon atoms adjacent to the vacancy with their lobes directed towards the vacancy site. With their bonding partners missing, they occur in the midgap region. At the same time, a localized state V$\pi$ gets introduced in the $\pi$ bands in the midgap region as well, the so called ``zero mode" state, whose energy is exactly zero in the nearest-neighbor tight binding approximation. These four states, localized around the vacancy center, can hold eight electrons in total taking into account the spin degeneracy. The level structure of the vacancy-induced states is shown in Fig. \ref{bandsplit}.

At the same time, electron counting arguments show that the vacancy releases four electrons to be occupied among the above localized states. These electrons include the
 three orphan $sp^2\sigma$  electrons, one from each of the three carbon atoms adjacent to the vacancy, plus one orphan $\pi$ electron,
 whose origin may be understood in the following way. Focusing on the $\pi$ states now and considering a vacancy on the $A$ sublattice,
the majority sublattice $B$ has one extra atom, $N_B-N_A =1 $, so that the total number of $\pi$ orbitals is $N_A+N_B$, which is the same as $2 N_A +1$.  Out of these, there is one zero-mode state and the electron-hole symmetry of the graphene lattice results in  $N_A$ band states below $E = 0$ and  the same number above it.
(See Fig. \ref{tbbands} for the $\pi$ band structure). So, of the $2 N_A +1$  $\pi$ electrons (one per atom),  $2 N_A $ fill up the lower bands, leaving a lone orphan $\pi$ electron.
 These four orphan electrons (three $\sigma$ and one $\pi$) occupy  the vacancy-induced states as indicated in Fig. \ref{sketchdos}.

The remaining sections are organized as follows. In Section II, we discuss the results of our DFT calculations. Section III discusses the crystal-field and Jahn-Teller splitting of the  vacancy-induced localized $\sigma$ states and Section IV is devoted to the vacancy-induced $\pi$ states. In Section IV A, we revisit the zero-mode theorem and in Section IV B, we present numerical results for the $\pi$ states from a numerical diagonalization of the tight-binding Hamiltonian before discussing the vacancy-induced $\pi$ states using the Green's function approach in Sections IV C and D. Finally the results are summarized  in Section V.

\section{\label{DFT} Density-Functional Calculations}

For the density-functional calculations, we used the all-electron spin-polarized linear augmented plane wave (LAPW) method\cite{lapw} with the general gradient approximation (GGA)\cite{perdew} for the exchange-correlation functional. A 72-atom $6 \times 6$ supercell was used which included one vacancy site.
 The LAPW basis functions included the carbon 2s and 2p valence orbitals and a momentum  cutoff of ${\it RK_{max}} = 5.2$ was used, with approximately 3500 basis functions and about 50,000 plane waves at each $k$ point. All atomic sphere radii were taken as 0.66 {\AA}. The maximum angular momentum for the wave function expansion inside the atomic sphere was kept at $l_{max}$ = 6. Thirty $k$ points in the irreducible Brillouin zone were found to be sufficient for converged results in the self-consistent calculations.

\begin{figure}
\centering
\includegraphics[width=4.0cm]{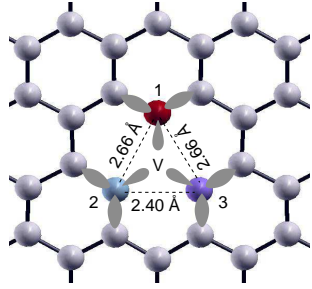}
\caption{
The  Jahn-Teller distorted planar carbon triangle obtained from the structural relaxation
using the all-electron spin-polarized LAPW-GGA method.
}
\label{relax}
\end{figure}

{\it Relaxed structure} -- First we performed a structural optimization of the lattice constant for pure graphene which yielded about the same lattice constant as the experimental value. For the vacancy calculation,  the lattice constant was held fixed at the experimental value and a structural relaxation was performed for the entire structure. The  optimization yielded a planar Jahn-Teller (JT) distorted
carbon triangle around the vacancy with the carbon atoms outside the triangle relaxed by a much smaller amount. For the carbon triangle, we found two long bonds of length 2.66 \AA\ each and a short bond of length 2.40 \AA\ (Fig. \ref{relax}), as compared to the 2.48 \AA\ for the undistorted structure. In terms of the
standard Jahn-Teller modes of the equilateral triangle, the magnitudes of the distortion are: $Q_0$ = 0.08 {\AA} (breathing mode), $Q_1$ = 0.166 {\AA} (symmetric bond-bending  mode), and $Q_2$ = 0 (asymmetric mode).\cite{Grosso}

The relaxed structure for the vacancy reported in the literature varies widely. While some have reported planar structures,\cite{Yazyev, Faccio, Singh, Yang, Forte, Choi} others have found non-planar structures with out-of-plane displacements varying from $\delta z \approx 0.12 - 0.47$  \AA.\cite{Briddon, nieminen, ma, Lim, Dai} We find that a paramagnetic relaxation (less accurate for the present problem) yields a non-planar structure $\delta z \approx 0.27$ \AA, while a spin-polarized calculation yields a planar structure, an observation made by
Faccio et al.\cite{Faccio} from their calculations as well using the SIESTA code.
We attribute this wide variation in the calculated relaxed structure in the literature partly  to the unusual nature of the V$\pi$ bound state, which falls off only as $1/r$, leading to a
larger width of the  V$\pi$ band in the supercell calculations than is expected from resonance broadening due to the $\pi$ band continuum.

The calculated vacancy formation energies agree much better between different calculations. Our result for
$E_V = E {\rm(graphene + vacancy)} - N^{-1}(N-1) E {\rm (graphene)}$, $N$ being the number of atoms in the graphene supercell, is 7.87 eV,  which compares well with the previous calculations\cite{Briddon, ma, Singh} of $ 7.4 - 7.8$ eV  as well as with the experimental value of $7.0 \pm 0.5 $ eV.\cite{Formation-Expt}

\begin{figure}
\centering
\includegraphics[width=11.0cm]{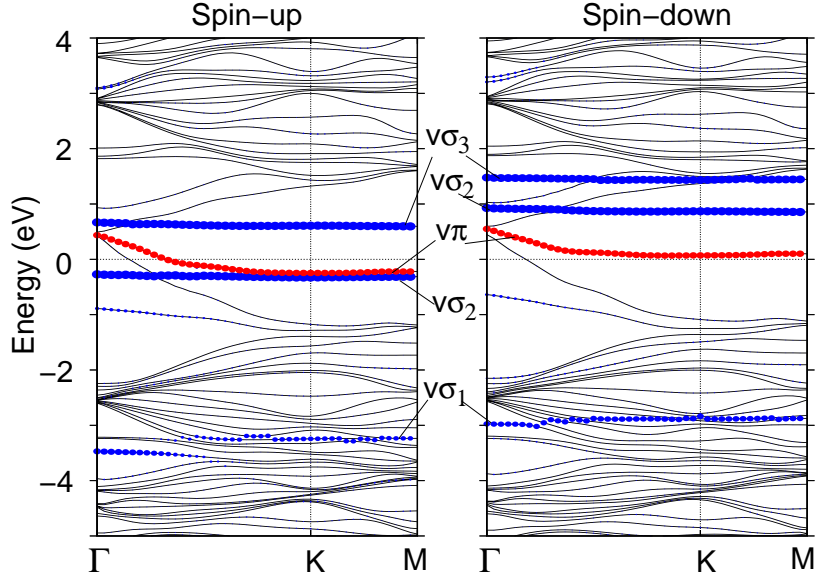}
\caption{ Spin polarized band structure of graphene  with a single vacancy in a 72-atom $ 6 \times 6$ supercell obtained from the density functional LAPW method. The vacancy induced  V$\sigma$ and V$\pi$ bands are indicated in blue and red, respectively.
Symmetry strictly forbids the admixture between $\sigma$ and  $\pi$ states for a planar relaxation around the vacancy, leading to flat V$\sigma$ bands (blue lines). The V$\pi$ bands are not flat owing to hybridization with the continuum $\pi$ states. The Dirac points $K$ and $K^\prime$ of the original graphene Brillouin zone get folded into the $\Gamma$ point of the supercell Brillouin zone. The zero of energy is taken to be the Fermi energy $E_F$.
}
\label{band}
\end{figure}

{\it Electronic structure} --
Fig. \ref{band} shows the band structure,
where the vacancy induced V$\sigma$ and V$\pi$ states are clearly seen. The momentum points in the Brillouin zone for the band structure plot are defined
as: $K =   \hat{x}/\sqrt{3}  +  \hat {y}   $ and
$M =   \hat {y}$ in units of $2 \pi 3^{-1} a^{-1}/n$ with $ n =6$ for the $6 \times 6$ supercell used in the calculation.
For this supercell, it can be easily seen by drawing both Brillouin zones that the Dirac points $K$ and $K'$ of graphene get folded into the $\Gamma$ point of the supercell Brillouin zone, so that remnants of the Dirac bands are seen at the $\Gamma$ point in Fig. \ref{band} just above $E_F$ (See also Fig. \ref{tbbands} for the folded graphene tight-binding $\pi$ bands for the same supercell and note the similarity between the tight-binding $\pi$ bands and the DFT bands, Fig. \ref{band}).
Due to symmetry, the $\sigma$ and  $\pi$ bands don't mix (strictly forbidden for the planar geometry, but also weakened significantly if the relaxed geometry is non-planar), which leads to clearly identifiable vacancy-induced V$\sigma$ bands.
The V$\sigma$ states originating from the dangling bonds are split due to the crystal field, Jahn-Teller, and exchange coupling as indicated
in Fig. \ref{bandsplit} and  discussed in more detail in Section \ref{sec:Vsigma}.
The dispersion of the V$\sigma$ bands in the band structure comes from the vacancy-vacancy interactions in different supercells or from the k-dependent interaction with the bonding and the anti-bonding  $\sigma$ bands,
both effects being small.
For non-planar relaxed structure, they should have a small resonance broadening due to the interaction with the $\pi$ band continuum.
Three electrons occupy these states leading to the occupation   V$\sigma_1 \uparrow \downarrow$, V$\sigma_2 \uparrow$,
with the remaining fourth electron occupying the V$\pi \uparrow$ state.

We now turn to a description of the effect of the  vacancy  on the $\pi$ states. Basically, the vacancy introduces a sharp, resonance state V$\pi$ in the midgap region.
The following summarizes the discussions in Section \ref{sec:Vpi}, which are important to keep in mind here: (i) If only NN tight-binding hoppings are kept, then the vacancy introduces a single localized state V$\pi$ at $E=0$ and of zero width called the zero-mode state, and its wave function decays as $\sim 1/r$ with distance in the linear-band approximation. (ii) Presence of the vacancy in each supercell does not affect the energy or the width of this state because of the result that the zero-mode wave function lives on the majority sublattice entirely and any changes in the minority sublattice does not affect it (in the supercell, all vacancies are located on the same, minority sublattice). (iii) However, due to the 2NN hopping as well as the exchange splitting,
the energy of V$\pi$ is different from zero, so that it now has a small but finite width due to resonance broadening   with the linear $\pi$ band continuum consistent with the STM experiments.\cite{Ugeda} (iv) In the supercell calculations, the V$\pi$ state acquires an extra broadening due to the slow $ 1/r$ decay of the V$\pi$ wave function,  because of the interaction between the supercells.

 {\it Dirac point} -- In Fig. \ref{band},  the Dirac point occurs above the $E_F$ (see the bands just above $E_F$ at the $\Gamma$ point, to which the standard Dirac points $K$ and $K^\prime$ have become folded).
 For the truly isolated vacancy, the location of the Dirac point above $E_F$ would mean that an infinite number of electrons  are transferred from the unfilled part of the Dirac cones to the lone vacancy site, which is impossible.
 Another way of seeing this is to consider first an infinite lattice without the vacancy. Obviously, the $E_F$ occurs at the Dirac point with zero density-of-states as usual. Now, if we introduce a single vacancy into the structure it can only affect the position of $E_F$ by $\sim 1/N$, where N is the total number of atoms in the lattice, so that $E_F$ remains unchanged for the infinite lattice. Of course, the electron states in the local neighborhood of the vacancy are modified, e. g., due to the resonance interaction with the vacancy states or due to the vacancy potential.  The Dirac-like bands seen just above $E_F$ at  $\Gamma$ in Fig. \ref{band} represent the effect of the vacancy on the electronic structure in the local neighborhood of the vacancy in the supercell calculation.

\begin{figure}
\centering
\includegraphics[width=6.5cm]{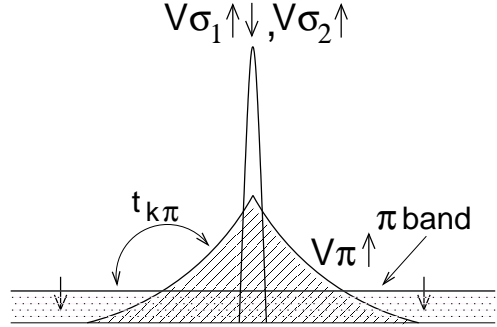}
\caption{Sketch of the  magnetic moment for an isolated vacancy, emphasizing the spatial extent of the various electronic states. The V$\sigma$ electrons are highly localized on the carbon triangle surrounding the vacancy, while the V$\pi$ electron is only ``quasi-localized" with its wave function decaying only  as $1/r$. Hund's-rule exchange aligns the V$\sigma$ and V$\pi$ electrons, producing a $S = 1$ state with the nominal magnetic moment of $ 2 \mu_B$.
This moment is however reduced by polarization of the $\pi$ band spins in the vicinity of the vacancy, described by an  antiferromagnetic Kondo-like coupling $t_{k\pi}$ between the $\pi$ bands and the localized V$\pi$ and  V$\sigma$ moments.
The $\pi$ band polarization is about $0.3 \mu_B$ in our DFT calculations, leading to the net magnetic moment of $1.7 \mu_B$.
}
\label{moment}
\end{figure}

 {\it Magnetic moment} -- The vacancy magnetic moment consists of two parts as shown schematically in Fig. \ref{moment}: (i) the localized moment coming from the vacancy states V$\pi$ and V$\sigma$ and (ii) the induced moment on the band electrons in the vicinity of the vacancy.     One can argue on general grounds that the first contribution should be  $2 \mu_B$ ($S = 1$), while the second contribution should reduce this value somewhat due to the antiferromagnetic Kondo-like coupling between the localized and the itinerant band spins.
Turning to the localized states,
the vacancy leaves four electrons to be occupied among the V$\sigma$ dangling bond states and the V$\pi$ zero-mode state.  Of these, three electrons will occupy the V$\sigma$ states, so that one electron resides on each of the three dangling bonds of the carbon triangle. The Coulomb interaction $U$ would prevent the occupation of a fourth V$\sigma$ state,
so that the remaining electron is energetically favored to occupy the $\pi$ states. The Hund's coupling between the V$\sigma$ and V$\pi$ electrons leads then to a $S=1$ state with a magnetic moment of $2 \mu_B$.  This basic picture is  illustrated in Figs. \ref{bandsplit} and \ref{moment} and it is  fully supported by the DFT bands, Fig. \ref{band}. This localized magnetic moment of $2 \mu_B$ is reduced due to the spin-polarization of the $\pi$ bands in the vicinity of the vacancy.

%

The spin polarization of the $\pi$ bands  can occur due to two factors: (i) the resonance coupling with the  V$\pi \uparrow$ electron with the $\pi$ continuum bands  and (ii) the Kondo-like antiferromagnetic interaction between the localized vacancy states and the continuum $\pi$ states. The first is  not well described in a supercell calculation due to the long-range nature of the V$\pi$ state and the second effect is intrinsically not well described within the band theory.

Our DFT calculations yield a magnetic moment of about $1.7 \mu_B$. This can be seen by estimating the number of holes in the small hole pocket in the two bands just above $E_F$ at the $\Gamma$ point  in the spin-up bands of Fig. \ref{band}. The spin-down bands must contain exactly the same number of  extra electrons missing from the spin-up bands. Without this pocket of holes, which represents the band polarization in the immediate neighborhood of the vacancy,  the magnetic moment would be exactly $2 \mu_B$, corresponding to the full occupancy of V$\sigma_1\uparrow \downarrow$, V$\sigma_2 \uparrow$, and V$\pi \uparrow$.
The existence of the hole pocket reduces this number.
We can estimate the number $n$ in the hole pocket by computing the total area of the two hole Fermi surfaces and comparing it to the area of the supercell Brillouin zone, which yields the value $n \approx 0.15$. Since the same number of electrons must be accommodated in the spin down bands, this would cause a net reduction of $N_\uparrow - N_\downarrow $ by $0.30$ leading to a net magnetic moment of $1.7 \mu_B$.

In the literature, the calculated magnetic moment varies widely, anywhere between $1.04 - 1.84 \mu_B$\cite{Yazyev, nieminen, Heggie, Briddon, ma,  Singh, Choi, Yang, Lim, Dai, Faccio, Forte, Palacios}. Typically, the lower values come from calculations, where the V$\pi \uparrow$ and V$\pi \downarrow$ bands overlap significantly. We suggest that the variation of the calculated magnetic moment in the literature is due to the intrinsic deficiency of the supercell method in estimating the $\pi$ magnetic moment due to the slow $1/r$ decay of the V$\pi$ state, which produces an extra broadening of the V$\pi$ state due to the supercell interaction and does not take into account the full anti-ferromagnetic polarization of the itinerant $\pi$ band states.



%
The exchange splitting  $\Delta$ of the V$\pi$ state is due to its overlap with the V$\sigma$ states which are localized on the three carbon atoms adjacent to the vacancy. It may be estimated from the expression
\begin{equation}
\Delta \equiv E (V\pi_\downarrow)  - E (V\pi_\uparrow) \approx J_H \times |\Psi_0 |^2 \approx 0.35 \ eV,
\end{equation}
where the Hund's-rule energy is typically $J_H \sim 0.9 - 1.0 $ eV  for the atoms and $|\Psi_0 |^2 \sim 0.4 $ is the combined total density of the V$\pi$ state on the carbon triangle as obtained from the DFT results.
The estimated exchange splitting is in agreement with the splitting seen in the DFT bands, Fig. \ref{band}.


{\it Relation to the Lieb's Theorem} --
The Lieb's theorem\cite{Lieb}  states that for the repulsive one-band Hubbard model on a bipartite lattice and half-filled band, the ground state has spin $S = (1/2) |N_B - N_A|$, $N_A$ ($N_B$) being the number of sites on the two sublattices. It is important to point out that the theorem  holds if we consider only the $\pi$-band system and also neglect the small second-neighbor interactions that couples the two sublattices. Thus, with a single vacancy present,  $|N_B - N_A| = 1$ so that according to the Lieb's Theorem we should have a net spin of $S=1/2$.
However, in addition to the $\pi$, we also have the $\sigma$ electrons. The Lieb result of $S=1/2$ for the $\pi$ electrons is now coupled to the spins of the three $\sigma$ electrons localized near the vacancy, leading to the net spin $S = 1$ as indicated in the summary figure, Fig. \ref{sketchdos}. We have already argued  that the magnetic moment of $2 \mu_B$ corresponding to $S = 1$ will be reduced due to the  polarization of the band electrons in the local neighborhood of the vacancy.

\section{\label{sec:Vsigma}  Vacancy-induced V$\sigma$ states}

The essential features of the density-functional results may be understood by simple tight-binding considerations of the effect of the vacancy on the $\sigma$ and the $\pi$ bands.
We study  the $\sigma$ states in this section followed by the  $\pi$ states in the next section.


The description of the vacancy-induced $ V \sigma$ states for graphene is rather simple.
In  graphene, the $sp^2\sigma$ states are removed away from $E_F$ due to strong interaction with neighbouring orbitals along the C-C bonds. However, with a vacancy present, the three $sp^2\sigma$ orbitals of the three NN carbon atoms with their lobes pointed towards the vacancy have their usual bonding partners missing, so that they occur near $E_F$, with their on-site energies $\epsilon_\sigma$ slightly below the $\pi$ orbital energies because of the $s$ orbital component present in the $\sigma$ states.

\begin{figure}
\centering
\includegraphics[width=6cm]{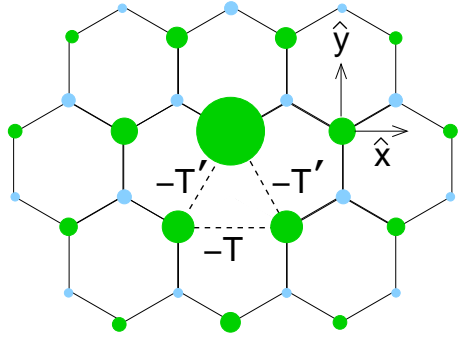}
\caption{ Spin density $n_\uparrow - n_\downarrow$ at different sites in graphene with a vacancy as obtained from the density-functional calculations.
Green (blue) circles indicate positive (negative) values, with
the area of the circle being proportional to the spin density. The spin moments on the carbon atoms other than the vacancy triangle are due to the $\pi$ electrons, which are long-ranged due to the slow $1/r$ decay of the V$\pi$ state. The hopping integrals $T$ and $T^\prime$  between the   $sp^2\sigma$ orbitals on the carbon triangle adjacent to  the vacancy has reference to the model discussed in Section \ref{sec:Vsigma}.
}
\label{spindens}
\end{figure}

The crystal-field splitting however will lift the three-fold degeneracy. The main feature can be described by taking into account the 2NN hopping  $T$ between the three dangling bonds in the undistorted triangle, leading to the $3 \times 3$ Hamiltonian:
\begin{eqnarray}
H_{cf} =\left( \begin{array}{ccc}
\epsilon_\sigma&-T&-T\\
-T&\epsilon_\sigma&-T\\
-T&-T&\epsilon_\sigma\\
\end{array}\right),
\label{HT}
\end{eqnarray}
diagonalization of which yields a double degenerate state at $E = T$ and a single degenerate state at $E = -2T$ as shown in Fig. \ref{bandsplit}, where we call this splitting the crystal-field splitting. The Jahn-Teller distortion of the triangle splits the double degenerate state further, which is described by the unequal hopping $ T \neq T^\prime$. Taking the isosceles-triangle relaxation found in our DFT results, two of the three hopping terms are modified into $T'$ as indicated in Fig. \ref{spindens}. From the DFT band structure, we find that  $T \approx 1.6 \ $ eV, while $T' \approx 1.2 \ $ eV.
The new eigenstates are: $E_{\sigma_1, \sigma_2} =  2^{-1} (-T \mp \sqrt{8T'^2 + T^2}) $ and $E_{\sigma_3} =  T$ with the corresponding (unnormalized) wave functions  $\Psi_{1,2} = ( (-T \pm \sqrt{8T'^2 + T^2})/ T', 1, 1)$ and $\Psi_3 = ( 0, -1, 1)$.
This simple model suggests a Jahn-Teller distortion of the carbon triangle surrounding the vacancy.

The Jahn-Teller interaction is of the type $E \otimes e$ (both electronic and vibrational states are doubly degenerate) in a trigonal (D$_{3h}$) symmetry. With this lattice distortion, the trigonal symmetry is broken.  The distortion removes the double degeneracy and the two states (shown in Fig. \ref{bandsplit} as $V\sigma_2$ and $V\sigma_3$) are now split  by the amount
$ 2^{-1} ( 3 T - \sqrt{8T'^2 + T^2} ) \approx 4 (T - T')/3 \approx 0.55 \ $ eV.
Since there are only three electrons available to the V$\sigma$ states, $V\sigma_1$ is occupied with two electrons, while the lone remaining electron occupies the $V\sigma_2$ state. The spin-degeneracy is removed by the Hund's coupling with the electron occupying the V$\pi$ state, producing the spin structure indicated in Fig. \ref{bandsplit}. The wave function $\Psi_2$ corresponding to the $V\sigma_2$ state shows that  the maximum weight ($\sim$ 66$\%$)  comes from the $sp^2\sigma$ dangling orbital of the apical atom of the carbon triangle, which is consistent with the spin density plotted in Fig. \ref{spindens}. The Jahn-Teller distortion is actually dynamic, with the carbon triangle tunneling between three equivalent minima on the adiabatic potential surface of the $E \otimes e$ Jahn-Teller problem, an issue we discuss elsewhere.\cite{DJT}

\section{Vacancy-induced V$\pi$ states}
\label{sec:Vpi}

In this Section, we discuss the origin of the localized state -- the so-called ``zero-mode" state  --  near the band center of the $\pi$ bands.  Understanding of the origin and the ``quasi-localized" nature of the  zero-mode state is an essential part of the interpretation of the full band calculation using the density-functional theory.

If only the NN interactions are present, the ``zero-mode" state is a localized state with energy exactly at the band center. If in addition the higher-neighbour interactions are also present but not too large, as is the case for graphene,\cite{brk-graphene}  then the localized state turns into a sharply-peaked resonance state owing to its overlap with the $\pi$ bands and occurs not too far from the band center.

\subsection{The existence of the zero mode state}
\label{subsec:zero-mode}
	
	According to the zero-mode theorem,\cite{castroneto} which  is in fact valid for any bipartite lattice with  NN interactions,  whenever there is an imbalance in the number of atoms  in the two sublattices in a bipartite lattice, viz., $ n = N_B - N_A > 0$, there are $n$ number of degenerate solutions with the eigenvalue $\epsilon_B$ (the on-site energy of the majority sublattice), with the wave functions residing entirely on this sublattice.  This can be seen from the following simple considerations as an alternative to Pereira et al.'s proof which used the rank-nullity theorem in linear algebra.\cite{castroneto}

We begin with the conjecture that there are some solutions where the wave functions  live completely on the majority sublattice ($B$) and proceed to find them. Thus we have
	\begin{eqnarray}
\left( \begin{array}{cc}
{\cal H}_{BB} & {\cal H}_{BA}\\
{\cal H}^{\dagger}_{BA} & {\cal H}_{AA}\\
\end{array}
\right)
\left (\begin{array}{c}
\Psi_B \\
0 \\
\end{array}
\right )
&=&
E\left( \begin{array} {c}
\Psi_B \\
0 \\
\end{array}
\right ),
\label{Hzero}
\end{eqnarray}
where $\Psi_B$ is a vector in the B sublattice of dimension $N_B$ and there is null contribution from the $A$ sublattice to the wave function.
It will be clear from the following discussion that for the theorem to hold, the B sublattice $N_B \times N_B$ Hamiltonian is restricted to the diagonal form
\begin{equation}
{\cal H}_{BB} = \epsilon_B I,
\label{HBB}
\end{equation}
and there are no restrictions
on the remaining parts of the Hamiltonian. The specific form of ${\cal H}_{BB}$ means that there is no site disorder, nor is there any interactions between the atoms on the B sublattice (hence it will fail if interactions beyond the NN are present, which will produce a non-diagonal ${\cal H}_{BB}$). However, such restrictions need not apply to the A sublattice, so that the $N_A \times N_A$ Hamiltonian ${\cal H}_{AA} $ for the minority sublattice can have diagonal disorder and also there is no restriction on the form of ${\cal H}_{BA}$ as well. This means that the A sublattice atoms can interact between themselves and with the B sublattice atoms as well without invalidating the theorem.

The wave function $\Psi_B$ thus satisfies
\begin{eqnarray}
{\cal H}_{BB} \Psi_B & = & E \Psi_B
\label{Hzero1}\\
{\cal H}_{BA}^\dagger \Psi_B & = & 0.
\label{Hzero2}
\end{eqnarray}
The first of these equations tells us that if the conjectured solutions of the form $(\Psi_B, 0)$ exist, then they must have the enegy $ E = \epsilon_B$ and there would be at most $N_B$ number of such degenerate solutions; the equation does not place any constraints on the individual components  of $\Psi_B$.

Turning to  Eq. \ref{Hzero2}, there are $N_B$ components of $\Psi_B$ to be determined but only $N_A < N_B$ equations are there to determine them. This means that the solutions can not be fully determined. However if we specify $N_B - N_A$ components of $\Psi_B$, then the remaining components are uniquely determined as linear functions of these components. These solutions are therefore of the form
\begin{equation}
\Psi_B = (\phi_1, \phi_2, ..., \phi_{N_B - N_A}; f_1, f_2, ..., f_{N_A}),
\label{form}
\end{equation}
 where we can choose the $\phi_i$'s arbitrarily and the  $f_i$'s  are then just linear combinations of $\phi_i$'s
($f_i = \sum_{j=1}^{N_B-N_A} c_{ij}\phi_j$, where the expansion coefficients
are determined by ${\cal H}_{BA}^\dagger$ in Eq. \ref{Hzero2}).
Thus the number of linearly independent solutions is given by the number of ways we can choose linearly independent solutions in the subspace
$(\phi_1, \phi_2, ..., \phi_{N_B - N_A})$, which is clearly
$N_B - N_A$. This proves the conjecture and the theorem.

It is easy to see why the theorem is not valid if there is on-site disorder on the majority sublattice or interactions beyond the NN, which introduces off diagonal terms in ${\cal H}_{BB}$. So, Eq. \ref{HBB} is not true anymore. This means that Eq. \ref{Hzero1} puts constraints on the components of $\Psi_B$ in order to satisfy the eigenvalue problem and as a result Eqs. \ref{Hzero1} and \ref{Hzero2} can not both be satisfied simultaneously.
For example, if we use the form Eq. \ref{form} which satisfies Eq.  \ref{Hzero2}, we are only left with the freedom to choose $\phi_1, \phi_2, ..., \phi_{N_B - N_A}$ and this is not enough to satisfy the eigenvalue problem of Eq. \ref{Hzero1}.
There is no such problem if $\cal{H}_{BB} = \varepsilon_B {\it I}$, since any vector $(\phi_1, \phi_2, ..., \phi_{N_B - N_A}; f_1, f_2, ..., f_{N_A})$ is a solution with $E = \varepsilon_B$.

	The theorem has an important bearing on the results of the supercell calculations, both tight-binding and density functional. In these calculations, the vacancies are repeated in each supercell, connected by the superlattice translational vectors, and hence are all located on the same sublattice, which forms the minority sublattice. If $n$ is the number of supercells in the crystal, then this is also the imbalance in the number of atoms in the two sublattices $n = N_B - N_A$. According to the theorem, there should be $n$ zero-modes in the Brillouin zone, which is also precisely the number of Bloch momentum points in the Brillouin zone. These states thus show up in the form of a dispersionless band in the tight-binding supercell calculations at $E = 0$.
	
\begin{figure}[h]
\centering
\includegraphics[width=12.0cm]{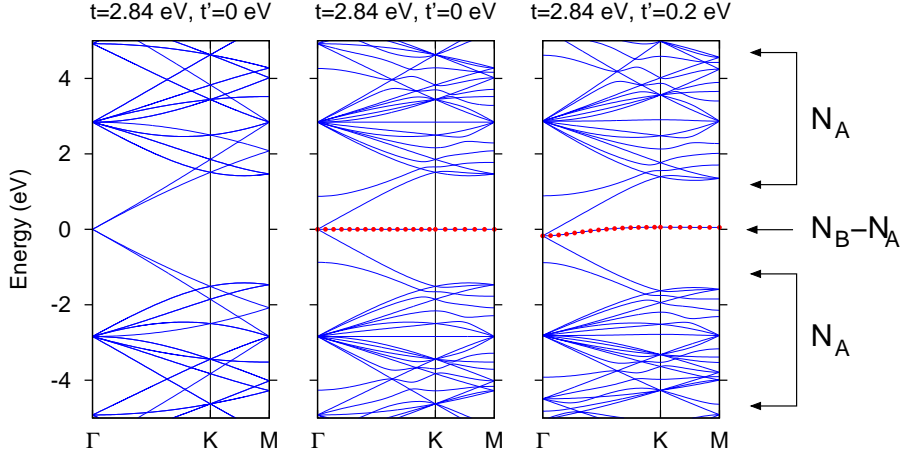}
\caption{ Tight-binding band structure obtained from the Hamiltonian Eq. \ref{TB} for the 72-atom $6 \times 6$ supercell both with  and without  a vacancy. The band structure without the vacancy is  shown in the left panel. The Dirac points $K$ and $K^\prime$ of the original graphene Brillouin zone get folded into the $\Gamma$ point of the supercell Brillouin zone and the two linear Dirac bands are clearly seen in the left panel.  The middle and the right panels show the zero-mode states (red dots) with and without the
second neighbour interaction $t^\prime$. In the NN tight-binding approximation (middle panel), all zero-modes have the same energy and live exclusively on the majority sublattice, while with the second neighbor interaction, the zero-mode states do have  a band dispersion and leak into the minority sublattice as well. The sublattice with the vacancy atoms is labelled
 $A$  and the total number of states  in different bands (not counting spin degeneracy) is shown on the right, with  $N_A$ and $N_B$ denoting the total number of atoms in the two sublattices  in the entire crystal.
}
\label{tbbands}
\end{figure}

If hopping beyond the NN is present or if the on-site energies of the different atoms are different, then the theorem does not hold. However, the hopping beyond the NN in graphene is small\cite{brk-graphene} and the on-site energies are negligibly different on sites close to the vacancy as obtained from the DFT calculations. Because these effects are small, a clearly identifiable, nearly-dispersionless zero-mode band is found in the DFT calculations as seen from Fig. \ref{band} as well as in the higher-neighbour tight-binding results [Fig. \ref{tbbands}], where the zero-mode band is marked by the red dots.
	

\subsection{Tight-binding results: Direct diagonalization of the Hamiltonian}
\label{sec:TB}

In order to further understand the formation of the zero-mode states,
we have studied the vacancy $\pi$ bands
with the standard tight-binding model of the $p_z$ orbitals containing both the nearest neighbour (NN) and the second- neighbour (2NN) interactions. In particular, we look for the vacancy-induced zero-mode states discussed in the previous Subsection.

The tight-binding Hamiltonian is
\begin{equation}
{\cal H}_{TB} = -t \sum_{\langle ij \rangle \sigma} c_{i \sigma} ^{\dagger}c_{j \sigma} + t^\prime\sum_{\langle\langle ij\rangle\rangle \sigma} c_{i \sigma}^{\dagger}c_{j \sigma} + H.c.,
\label{TB}
\end{equation}
where $-t$ and $t'$ are the NN and the 2NN interactions with the signs chosen such that  $ t, t' > 0$ ($t \approx 2.91$ eV and $t' \approx 0.16 $ eV for graphene\cite{brk-graphene}).

\begin{figure}
\centering
\includegraphics[width=12.5cm]{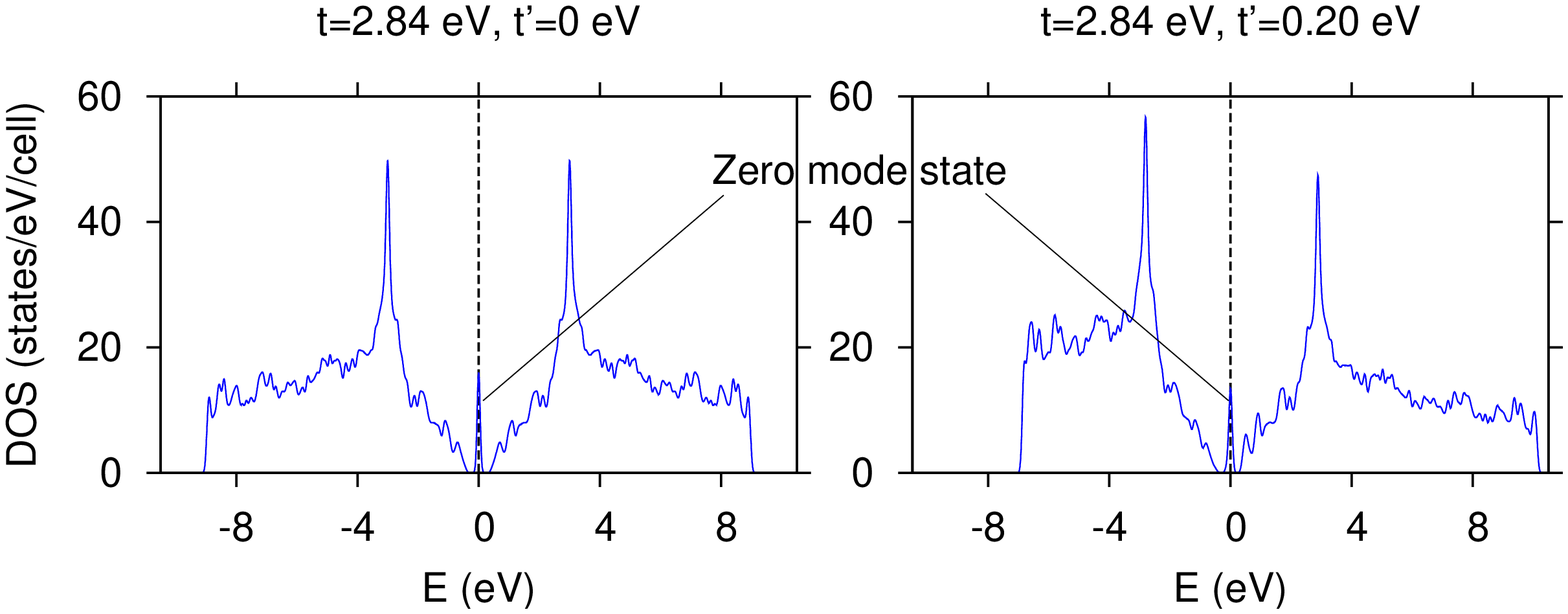}
\caption{\label{tbdos}The tight-binding $\pi$ densities of states of graphene with a vacancy  with NN interacations only (left) and with both the NN and the 2NN interactions present (right) as obtained from the  tight-binding Hamiltonian Eq. \ref{TB}. }
\end{figure}

The band structures and the densities-of-states   are shown in Figs. \ref{tbbands} and \ref{tbdos}. The electron counting in the band structure Fig. \ref{tbbands} is as follows. Both the lower and the upper bands contain in total (integrated over the Brillouin zone) $N_A$ states each, while the zero-mode band contains $N_B - N_A$ states, making a total of $N_A + N_ B$ states, as it must be the case. We have one $\pi$ electron per site present in the system, so that taking spin into account, the entire lower subband is full, while the zero-mode states are half full.
For the single vacancy ($N_B - N_A = 1$), this leads to a single occupied electron in the zero-mode states, resulting in $ S = 1/2$
 in agreement with the Lieb's Theorem\cite{Lieb}.

As discussed in the previous Section,  if only the NN interactions are present, we should have $N_B - N_A$ number of zero-mode states at $E=0$ exactly. This is why the zero-mode band in the middle panel of Fig. \ref{tbbands} is completely flat. However, if the 2NN interactions are also present, then the energies of the zero-mode states are not guaranteed to be the same and we see a spread in the energy of these states, which shows up as a dispersion in the zero-mode band, as seen in the right panel of Fig. \ref{tbbands}.

Here, the vacancy site was modeled by simply removing a lattice site, corresponding to the vacancy potential $U_0 = \infty$. In a real material, however, $U_0$ is large but finite. The effect of a finite $U_0$ is that (a) It causes the zero-mode state to occur slightly below the mid-gap ($E=0$) and (b) The sharp zero-mode state turns into a resonance state due to interaction with the continuum $\pi$ bands. This is best described with the Green's function approach discussed in the next Subsection.

\subsection{Impurity Green's Function and the zero-mode state in the $\pi$ bands}

In this Section, we study the effect of a single impurity on the $\pi$ electron states by studying the Dyson's equation and show the emergence of the zero-mode state as the strength of the impurity potential $U_0$ is gradually increased. For the vacancy, this potential is large but finite, so that the results obtained in this Section are helpful in understanding the nature of the zero-mode state in the actual structure with a finite vacancy potential.

The wave function of the zero-mode state is obtained from the Lippmann-Schwinger equation. Consistent with the previous results,\cite{castroneto,Pereira2} we find that (a) The zero-mode state consists of wave functions from the majority sublattice only  and (b) It is quasi-localized decaying as $1/r$ as a function of distance from the vacancy in the limit of the linear-band approximation. In addition to these known results, our analysis allows us to (a) obtain the oscillatory phase factors in the zero-mode wave function due to the interference of the two Dirac points  and (b) compare the linear-dispersion results with the full tight-binding band result by computing the GFs for large distances in both cases.

\begin{figure}
\centering
\includegraphics[width=8.0cm]{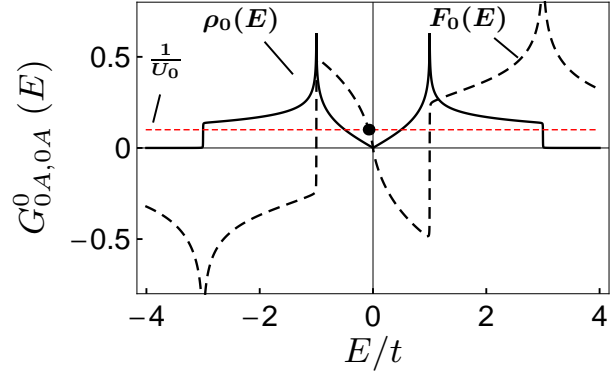}
\caption{On-site GF $G^0_{0A, 0A}(E)$ for the $\pi$ bands computed from the Horiguchi method and the energy of the resonance state, indicated by the black dot, obtained from the Dyson's equation: $U_0 F_0(E) =1$. As $U_0 \rightarrow \infty$, the solution moves to $E \rightarrow 0$ leading to the sharply-localized zero-mode state at the band center.}
\label{Fig-Dyson}
\end{figure}

\begin{figure} [h]
\centering
\includegraphics[width=6.5cm]{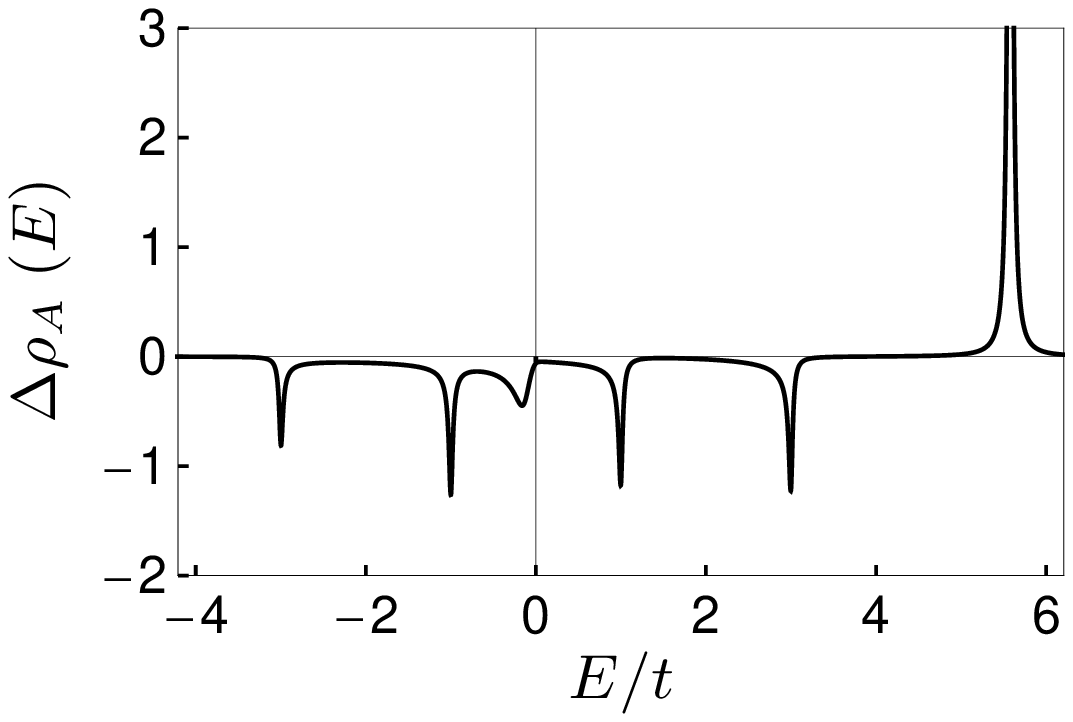}
\includegraphics[width=6.5cm]{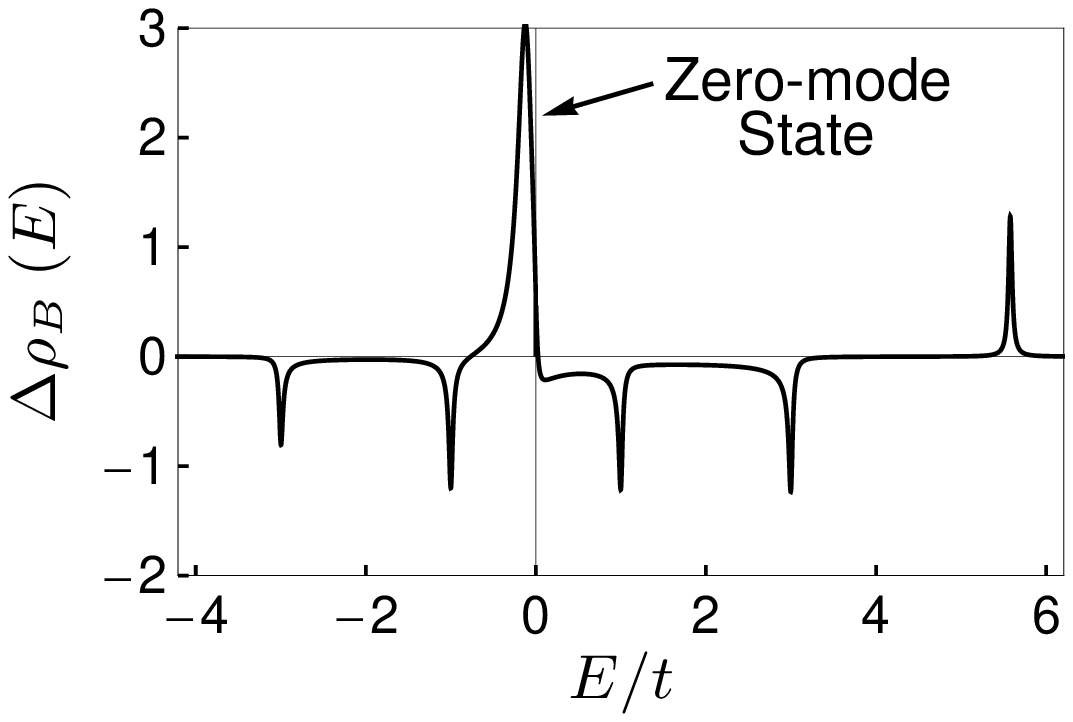}
\caption{Change in the total sublattice DOS  due to the addition of the impurity
with $U_0 = 5 t$ as computed from  the factors multiplying the coefficient ``$1/N$" in Eqs. \ref{DOSA} and \ref{DOSB}. The emergence of the zero-mode state on the $B$ sublattice  at $E = 0$ is clearly seen, which grows into a $\delta$ function as $U_0 \rightarrow \infty$.  The remaining changes in the DOS go to zero as $1/N$, except for the (unimportant) bound state
on the $A$ sublattice (occurring at $E/t \sim 6$ in the top figure), which becomes a $\delta$-function bound state with $E \rightarrow \infty$  in the limit of $U_0 \rightarrow \infty$.}
\label{Delta-Rho}
\end{figure}

The vacancy is modeled by  adding an on-site perturbation $V$ to  the unperturbed NN (NN) tight-binding (TB) Hamiltonian, so that
\begin{equation}
{\cal H}  =
{\cal H}_0+ V,
\label{hamil}
\end{equation}
where ${\cal H}_0=-t \sum c^{\dagger}_{i\alpha}c_{j\beta} + H. c.$, $i\alpha$ being the site-sublattice index, and
\begin{equation}
 V=U_0 c^{\dagger}_{0A}c_{0A},
\label{imp}
\end{equation}
 where $U_0$ is the strength of the potential due to the impurity on the $A$-sublattice in the central cell. The vacancy corresponds to the value of $U_0 \rightarrow \infty$.

The key quantity of interest here is the unperturbed GF,  $G^0(E) = (E+i \eta - {\cal H}_0)^{-1}$, the calculation of which we have discussed in our earlier paper where we studied the RKKY interaction in graphene.\cite{Sherafati} As usual, the imaginary part of the GF contains the information about the density-of-states: $\rho^0(E)= -\pi^{-1} {\rm Im} \  G^0 (E)$. The GF $G(E)$ in the presence of the perturbation  will be obtained from the Dyson's equation.

Since we will be interested in the local density-of-states (LDOS) on the various carbon sites and how they are modified by the presence of the impurity, we will need to calculate the real-space matrix elements $G^0_{i\alpha, j\beta} (E) \equiv \langle i\alpha| G^0(E)|j\beta\rangle$. This may be done by going to the momentum space and defining the Bloch functions for the electrons $|\vec{k}\alpha\rangle= N^{-1/2} \sum_i  e^ {i\vec{k}.\vec{r}_{i \alpha}} |i \alpha\rangle$ with $\vec{r}_{i \alpha}=\vec{R}_i+\vec{\tau}_\alpha$ being the position vector of the $\alpha$-{th} atom in the $i$-{th} unit cell.
The unperturbed Hamiltonian ${\cal H}_0$  in this basis set becomes ${\cal H}_{\vec{k}}=
 \left( \begin{array}{cc}
0 &  f(\vec{k})\\
f^*(\vec{k}) & 0
\end{array}\right)$,
where $f(\vec{k}) = -t \ (e^{i  \vec{k} \cdot  \vec{d}_1} + e^{i  \vec{k} \cdot  \vec{d}_2} + e^{i  \vec{k} \cdot  \vec{d}_3} ) $  and $\vec{d}_1$, $\vec{d}_2$, and $\vec{d}_3$ are the positions of the three nearest neighbors. Diagonalization of the Hamiltonian yields the eigenenergies $E(\vec{k}) = \pm |f(\vec{k})|$, which when expanded around the Dirac points lead to the usual linear band structure $E(\vec{q}) = \pm v_F |\vec{q}|$, where $\vec{q} = \vec{k} - \vec{K}_D$ is the deviation from the Dirac point in the Brillouin zone. Here the Fermi velocity $v_F = 3 t a /2$, where '$a$' is the carbon-carbon bond length. Note that unlike our previous work,\cite{Sherafati} $v_F$ here is defined to be a positive quantity, since `$t$' is positive.

The real-space GFs are conveniently obtained by first calculating the momentum-space GF, which can be easily shown to be
$
G^0 (\vec{k}E)  \equiv \langle \vec{k} \alpha | G^0 (E) | \vec{k}\beta\rangle =  (E+i \eta + {\cal H}_{\vec{k}}) ((E+i \eta )^2 - |f(\vec{k})|^2)^{-1}$. A Fourier transform then yields the
real-space unperturbed GF, viz.,
\begin{equation}
G^0_{i\alpha, j \beta} ( E) = \frac{1}{N}\sum_{\vec{k}} e^{i \vec{k} \cdot (\vec{r}_{i \alpha}-\vec{r}_{j \beta})}   G^0_{\alpha \beta} (\vec{k} E),
\label{GF-Site}
\end{equation}
which can be calculated by simply a brute-force summation over the Brillouin zone. It can also be computed by a second method using the Horiguchi recursive technique.\cite{Horiguchi, Sherafati} However, the latter, although computationally fast, has convergence problems\cite{Berciu}  for distances $ |\vec{R}_i - \vec{R}_j|  \ge 7 a$ or so, so that this is a better method to use only for smaller distances.

    The perturbed GF is related to the unperturbed GF through the Dyson's equation: $G = G^0+ G^0 V G$.   Using the localized form of the impurity potential, Eq. \ref{imp}, and taking the matrix elements, we immediately get for the real-space GF, the result
 \begin{equation}
G_{i\alpha, j \beta}(E)=G^0_{i\alpha, j \beta}(E)+U_0\times
              G^0_{i\alpha, 0 A}(E)G_{0A, j \beta}(E).
\label{Dyson}
\end{equation}

We are specifically interested in the on-site GFs with
$\alpha=\beta$ and $R_i=R_j$, which give the LDOS  on the $A$ and $B$ sites at distance $r_{i\alpha} = R_i +\tau_\alpha$ from the impurity site. Eq. (\ref{Dyson}) is easily inverted to yield the perturbed $G(E)$ in terms of the unperturbed $G^0(E)$, viz.,

 \begin{equation}
G_{i\alpha, i\alpha}(E)=G^0_{i\alpha, i\alpha}(E) + \frac{U_0 G^0_{i\alpha,0A}(E)  G^0_{0A, i\alpha} (E)}  {1-U_0G^0_{0A, 0 A}(E)}.
\label{GAA}
\end{equation}

The LDOS at different sites may be obtained by taking the imaginary parts of the diagonal elements of the GF:
$
\rho_{i\alpha} (E) = - \pi^{-1} {\rm Im} \ G_{i\alpha, i\alpha}(E).
$
It immediately follows from Eq. \ref{GAA} that the LDOS at the impurity site has the much simpler form
\begin{equation}
\rho_{0A }(E)= \frac{\rho_0(E)}{(1-U_0 F_0(E))^2+(\pi U_0\rho_0(E))^2},
\label{LDOSAA00}
\end{equation}
where $\rho_0 (E) = -\pi^{-1} {\rm Im} \ G^0_{0A, 0A}(E) $ is the unperturbed LDOS at the central site, which is of course the same for every site in unperturbed graphene, and $F_0(E)= {\rm Re} \  G^0_{0A, 0A}(E)$.
Note that we have defined here $\rho_0 (E)$ to be the unperturbed density-of-state per sublattice per spin (which is independent of the sublattice or the cell index) and $\rho_{i \alpha }(E)$ is the corresponding perturbed quantity for the $i \alpha$ site.

The resonance condition follows from Eq. \ref{LDOSAA00}, viz.,
\begin{equation}
1-U_0F_0(E_0)=0,
\label{E0}
\end{equation}
where $E_0$ is the resonance energy.
The graphical solution of this equation  is shown in Fig. \ref{Fig-Dyson}. There are four solutions for $E_0$: The two solutions at $E_0 = \pm t$ do not produce much change in the DOS, as may be inferred from Eq. \ref{LDOSAA00}, due to the diverging density-of-states $\rho_0 (E)$ there, and the bound state with $E_0 \rightarrow U_0$ for large $U_0$ is inconsequential because it is removed to $\infty$, which then leaves the sole resonance state indicated by the black dot in Fig. \ref{Fig-Dyson}. Its energy goes to zero in the limit $U_0 \rightarrow \infty$ and the oscillator strength to one, producing the zero-mode state for the vacancy.

The total DOS in the presence of the perturbation may be computed by taking the trace of Eq. (\ref{GAA}) for the entire lattice. Using the identity
$\sum_{i \alpha} G^0_{i\alpha, 0 A} (E) G^0_{0A, i \alpha} (E)=-dG^0_{0A, 0A}(E) / dE$ and some tedious algebra, the result is

\begin{eqnarray}
\fl
\rho_\textrm{tot}(E)=2\rho_0(E)+\frac{1}{N} \times
\frac{- U_0 [U_0 \rho_0(E)F'_0(E)+\rho'_0(E)(1-U_0 F_0(E))]}
 {(1-U_0 F_0(E))^2+(\pi U_0\rho_0(E))^2}.
\label{DOSTot}
\end{eqnarray}
Similarly, by taking the trace of Eq. (\ref{GAA}) over the cell index only, the individual sublattice DOS may be found, which for the $A$ sublattice reads as
\begin{equation}
\fl
\rho_{A}(E)=\rho_0(E) + \frac{1}{N} \times \frac{-U_0 [(1-U_0 F^0(E)) \ {\rm Im} \xi(E)-\pi U_0 \rho_0(E) \ {\rm Re}\xi(E)]} {\pi[{(1-U_0 F_0(E))^2+(\pi U_0\rho_0(E))^2}]},
\label{DOSA}
\end{equation}
where $\xi(E)=(1/N)\sum_k[G^0_{AA}(kE)]^2$ and the densities of states are, again, per sublattice and per spin.
A similar expression for $\rho_B (E)$ reads
\begin{eqnarray}
\fl
\rho_{B}(E)&=&
\rho_0(E)   + \frac{1}{N}  \nonumber
\\ \fl
&\times&    \frac{-U_0 [(1-U_0 F^0(E))(\pi \rho'_0(E)-Im\xi(E))+\pi U_0 \rho_0(E)(F'_0(E)+Re\xi(E))]
} {\pi[{(1-U_0 F_0(E))^2+(\pi U_0\rho_0(E))^2}]}.
\label{DOSB}
\end{eqnarray}
%
It can be verified from Eqs. (\ref{DOSTot}) - (\ref{DOSB}) that
$
\rho_\textrm{tot}(E) = \rho_{A}(E) + \rho_{B}(E),
$
so that these equations are consistent.

\begin{figure}
\centering
\includegraphics[width=7.0cm]{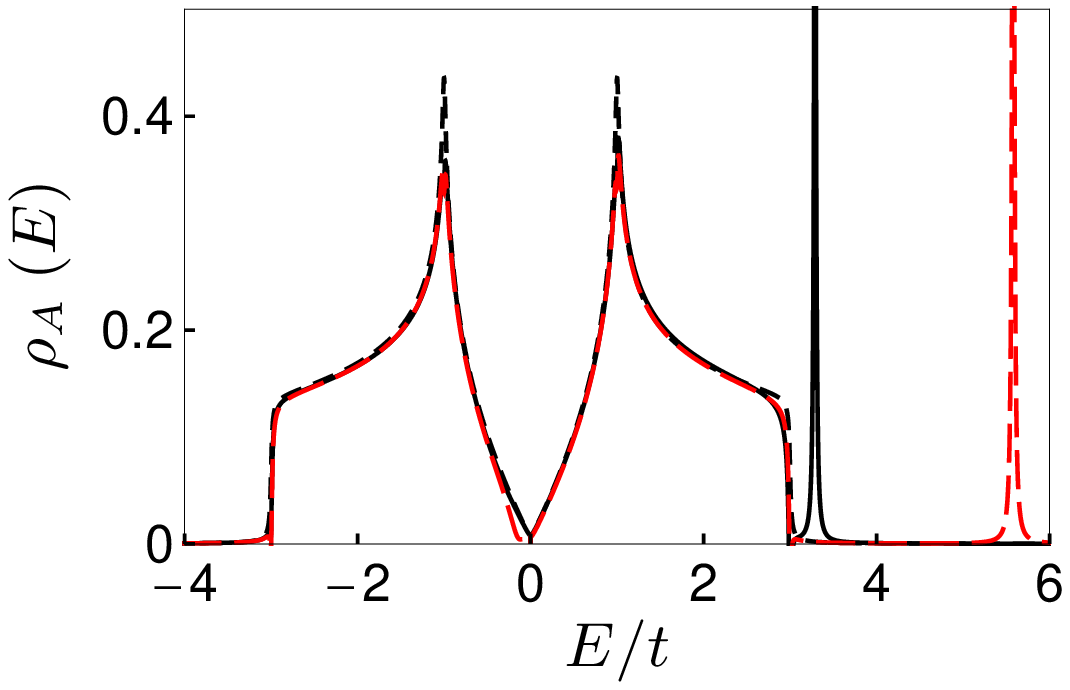}
\includegraphics[width=7.0cm]{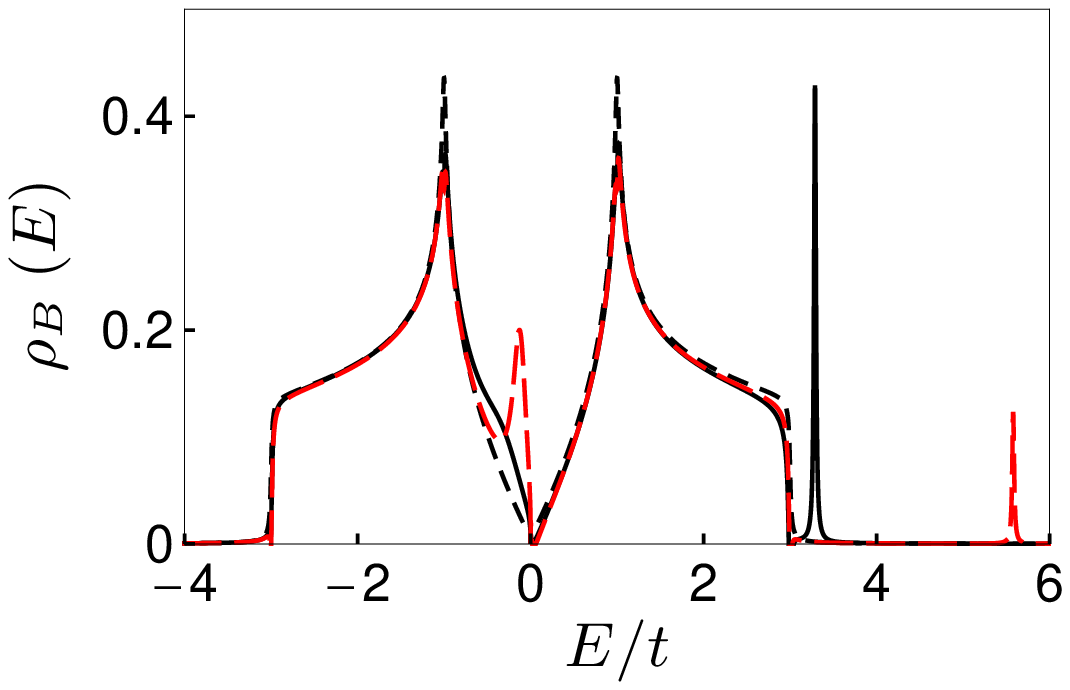}
\caption{ Total density-of-states for the $A$ sublattice $\rho_A(E)$ (top) and the  $B$ sublattice $\rho_B(E)$ (bottom) for different values of the impurity potential $U_0 / t =0, 2,  {\rm and\ } 5$, denoted by the black dashed, black solid, and red dashed lines respectively. These results are obtained from Eqs. \ref{DOSA} and \ref{DOSB} by using $N = 20$ for the purpose of plotting.
The figure shows the evolution of the zero-mode state at $E = 0$, which lives completely on the $B$ sublattice in the limit of $U_0 \rightarrow \infty$, i.e., opposite to the sublattice in which the vacancy is introduced.
}
\label{TDOS}
\end{figure}

\begin{figure}
\centering
\includegraphics[width=6.5cm]{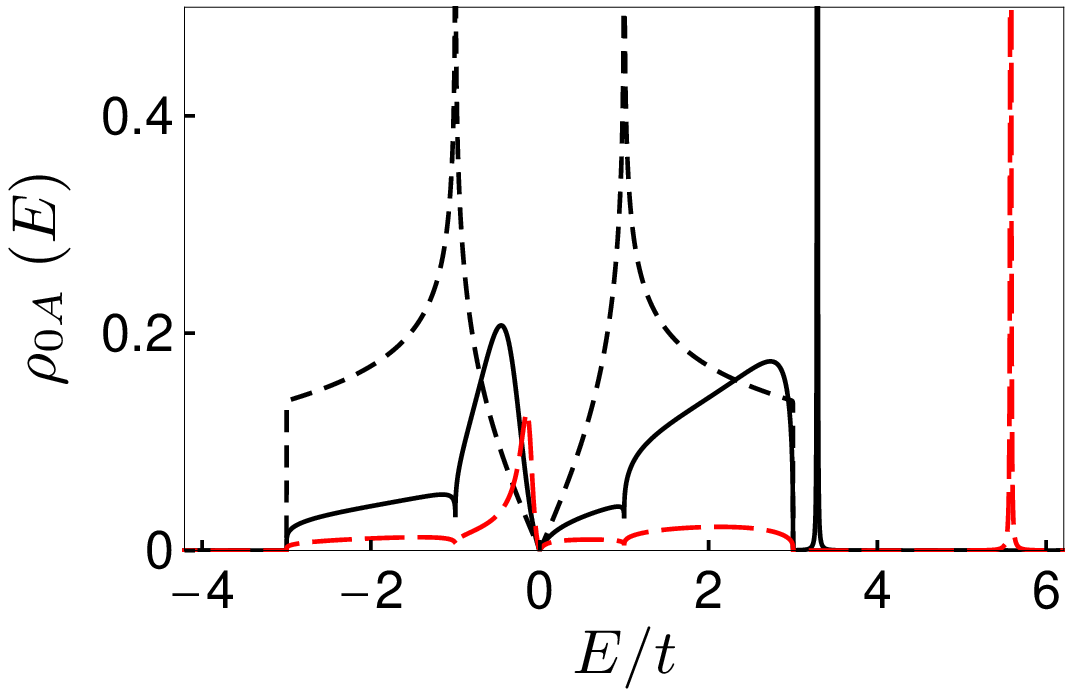}
\includegraphics[width=6.5cm]{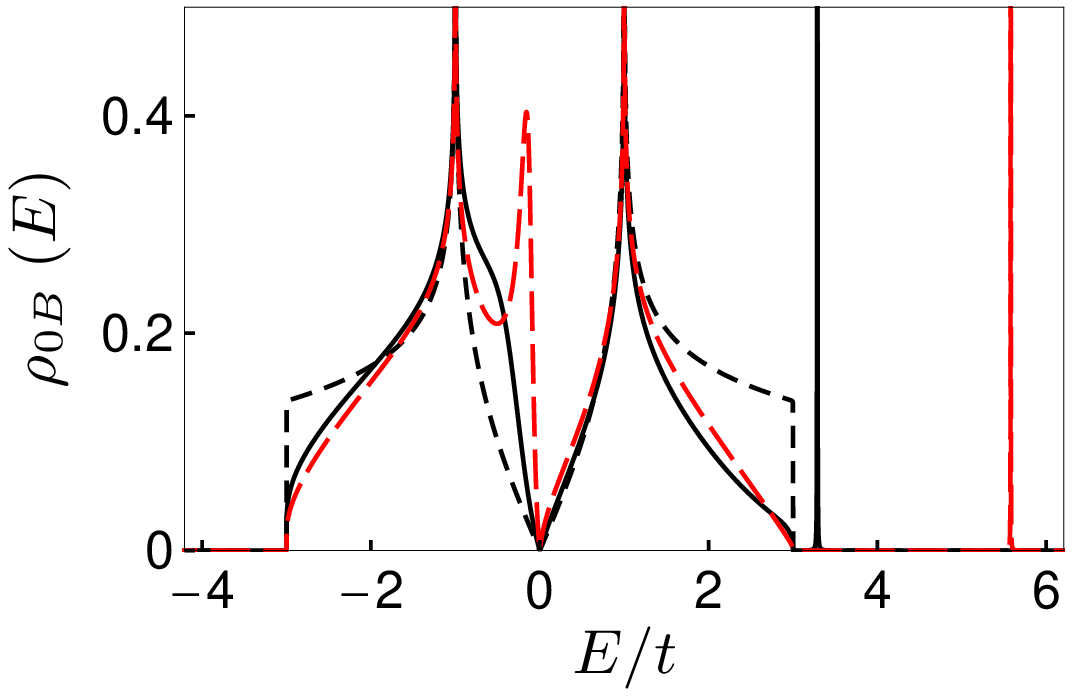}
\includegraphics[width=6.5cm]{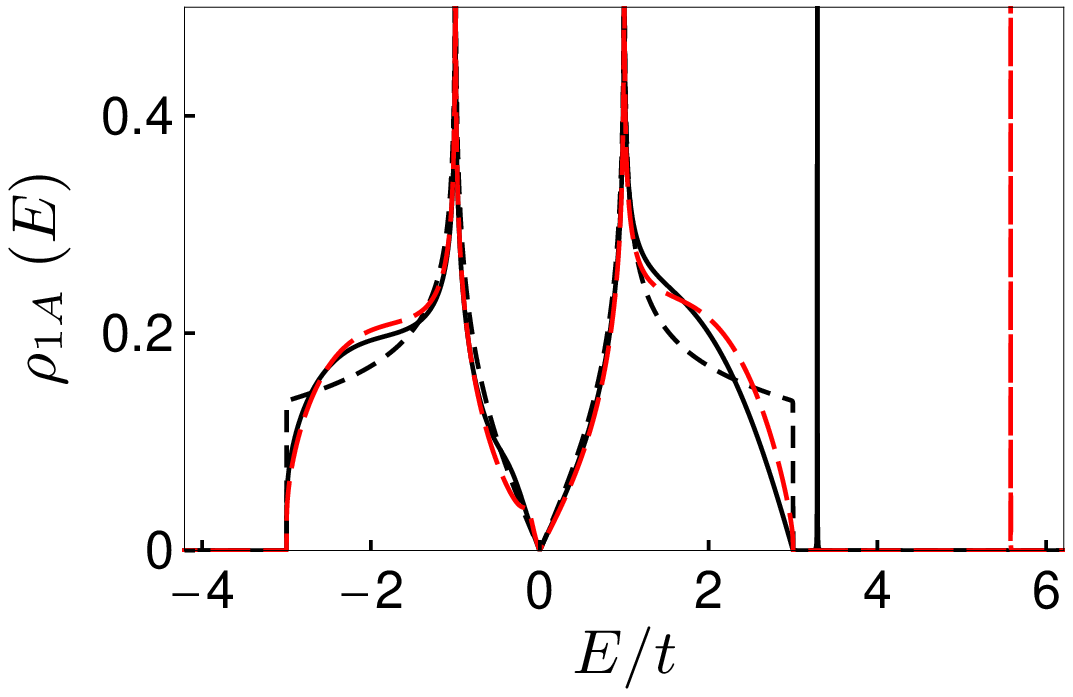}
\caption{ Local density-of-states at the impurity site $\rho_{1A} $ (top), the NN site $\rho_{0B} $ (middle), and the next NN site $\rho_{1A}$ (bottom) obtained from Eqs. \ref{GAA} and \ref{LDOSAA00} for different strengths of the impurity potential  $U_0 / t = 0, 2, {\rm and} \ 5$, denoted by black dashed, black solid, and red dashed lines respectively.
As $U_0 \rightarrow \infty$, the top LDOS goes to zero (except for the bound state beyond the top of the band whose energy goes to $\infty$), and the zero-mode state lives only on the $B$ sublattice, as indicated from the middle and the bottom panels.
The prominent zero-mode peak in the middle panel for $U_0/t = 5$ will develop into a $\delta$-function peak at $E=0$ as the impurity potential $U_0 \rightarrow \infty$.
}
\label{LDOS}
\end{figure}

The numerical results are summarized in Figs. \ref{Delta-Rho}, \ref{TDOS}, and \ref{LDOS}. The factors multiplying the $1/N$ in  Eqs. \ref{DOSA} and \ref{DOSB} are the changes in the DOS $\Delta \rho_A(E)$ and $\Delta \rho_B(E)$ introduced by the impurity potential, which are shown in Fig. \ref{Delta-Rho}. Fig. \ref{TDOS} shows the emergence of the zero-mode in the density-of-states with $E=0$ and that this state resides entirely on the $B$ sublattice in the limit $U_0 \rightarrow \infty$. Fig. \ref{LDOS} shows the LDOS on the impurity site ($\rho_{0A}$)
and on the nearest ($\rho_{0B}$) and the next nearest sites ($\rho_{1A}$).

\begin{figure}
\centering
\includegraphics[width=6.0cm]{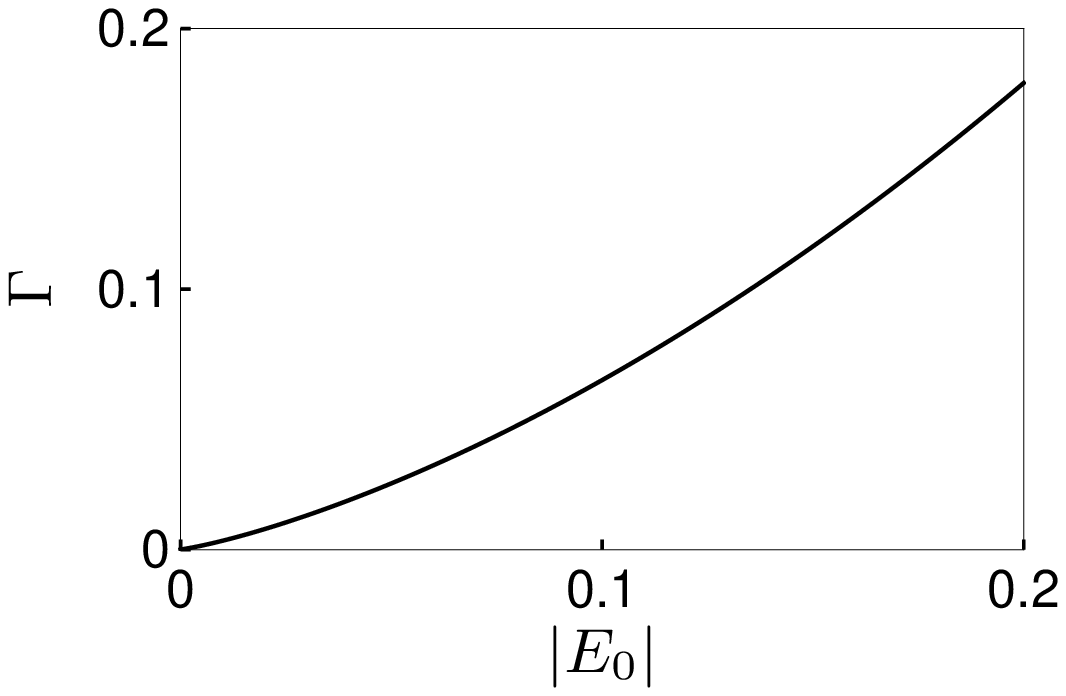}
\caption{ Resonance width of the zero-mode state (V$\pi$) as a function of the resonance energy $E_0$. Both $E_0$ and $\Gamma$ are in units of the NN hopping, with the value $t \approx 2.56$ eV \cite{brk-graphene}
 if only the NN hopping is kept.}
\label{Width}
\end{figure}

The width of the resonance peak increases with the resonance energy $E_0$. Keeping the linear term in the expansion of $F_0(E)$, viz., $F_0(E) \approx U_0^{-1}+F'_0(E_0)(E-E_0)$, Eq. (\ref{DOSTot}) yields the Lorenzian lineshape
\begin{equation}
\rho_{\textrm{tot}}(E) \approx 2\rho_0(E)+\frac{1}{\pi N}\frac{\Gamma}{(E-E_0)^2+\Gamma^2},
\label{Lorentz}
\end{equation}
where the resonance width is given by
\begin{equation}
\Gamma=- \pi \rho_0(E_0)/ F'_0(E_0).
\end{equation}
The width is zero if $E_0 = 0$ and increases with energy as shown in Fig. (\ref{Width}).

\subsection{Impurity state wave function}

In this section, we study the impurity state V$\pi$ wave function
 from the Lippmann-Schwinger equation. The analysis allows us to obtain the well known $1/r$ decay of the vacancy state; however, in addition we also obtain the oscillatory behavior of the wave function due to the interference between the two Dirac cones.

 Our starting point is the Lippmann-Schwinger equation
 $| \Psi \rangle = | \Psi^0 \rangle  + G^0 V | \Psi \rangle$, where $| \Psi^0 \rangle$ is the unperturbed state.
    For the localized impurity potential on the central $A$ site, $V = U_0 |0A \rangle \langle 0A|$, the Lippmann-Schwinger equation leads to the wave function
\begin{equation}
\Psi_{i \alpha} \equiv \langle i \alpha | \Psi\rangle = \Psi_{i \alpha}^0 +
 \frac{U_0 G^0_{i\alpha , 0A} (E) \Psi_{0 A}^0}{1- U_0 G^0_{0A , 0A} (E)}.
 \label{Psi-LS}
\end{equation}
 We are interested in the low-energy behavior, since that's the region where the resonance state gets introduced by the impurity as seen from Fig. \ref{Fig-Dyson}. The GFs for the full tight-binding band structure as well as for the linear bands were computed in our previous work.\cite{Sherafati}  For the linear bands and in the low energy limit, the results are
 \begin{eqnarray}
  G^0_{i A, 0A} (E) &=& - \beta \frac{A_c E}{2 \pi v_F^2} K_0 (\frac{-i E r}{v_F}),  \nonumber \\
G^0_{i B, 0A} (E) &=& \alpha \frac{A_c E}{2 \pi v_F^2} K_1 (\frac{-i E r}{v_F}),
\label{LS-GF}
 \end{eqnarray}
 where $A_c$ is the unit cell area in graphene, $K_0$ and $K_1$ are the modified Bessel functions of the second kind and
 $\vec r $ is the distance vector between the two atoms:  $\vec r = \vec {r}_{iA} - \vec {r}_{0A}$ for the first GF and $\vec r = \vec {r}_{iB} - \vec {r}_{0A}$ for the second.
 The multiplicative factors are
  $ \beta =  e^{i \vec{K} \cdot \vec{r} } +  e^{i \vec{K}^\prime \cdot \vec{r} }$, which is a real number for the graphene lattice and
 \begin{equation}
  \alpha =  e^{-i \pi/3} ( e^{i (\vec{K} \cdot \vec{r} - \theta_r)} -  e^{i (\vec{K}^\prime \cdot \vec{r} + \theta_r)}),
\label{alpha}
\end{equation}
  which is purely imaginary
and the polar angle $\theta_r = \tan^{-1} (y/x)$  is defined with the $x$ direction taken to be along the vector $\vec{K'}-\vec{K}$, which are two adjacent Dirac points in the Brillouin zone.

    Using the small $z$ expansion for the Bessel functions:\cite{GRHandbook}
    $K_0 (z) = - \ln (z/2) - \gamma$ and $K_1(z) = z^{-1} + 2^{-1} z \ln (z/2) + (\gamma - 1/2) z / 2$, where $\gamma\approx 0.577$ is the Euler-Mascheroni constant, we find the Bessel functions entering the expressions for the GFs (Eq. \ref{LS-GF})  to become, in the low energy limit:
\begin{eqnarray}
 K_0 (-\frac {iEr}{v_F})  &=& \frac{i\pi}{2} {\rm sign} (E) + \ln \frac{2v_F}{|E|r} - \gamma,     \\
K_1 (-\frac {iEr}{v_F})  &=& -\frac{v_F}{iEr} -  \frac {iEr} { 2 v_F}  \ln (-\frac {iEr} { 2 v_F})
 -(\gamma - \frac{1}{2}) \frac {iEr} { 2 v_F}.    \nonumber
\end{eqnarray}
Plugging these into Eq. \ref{LS-GF}, taking the $E \rightarrow 0$ limit , and retaining the lowest-order terms in energy, we find the results
\begin{eqnarray}
G^0_{i A, 0A} (E) &=&\frac{A_c  \beta }{2 \pi v_F^2} \times ( E \ln \frac{|E|r}{2 v_F} + \gamma E - \frac{i\pi}{2} |E| ),   \nonumber  \\
G^0_{i B, 0A} (E) &=&  \frac{A_c {\rm Im} \alpha}  {2 \pi v_F^2} [  -\frac{v_F}{r} + \frac{E^2r}{2v_F} (\gamma-2^{-1} +    \ln \frac{|E|r}{2 v_F}   )   \nonumber  \\
& - &      \frac{i r \pi}{4 v_F} E^2 {\rm sign} E ].
\label{GF-lowE}
\end{eqnarray}

\begin{figure}
\centering
\includegraphics[width=7.0cm]{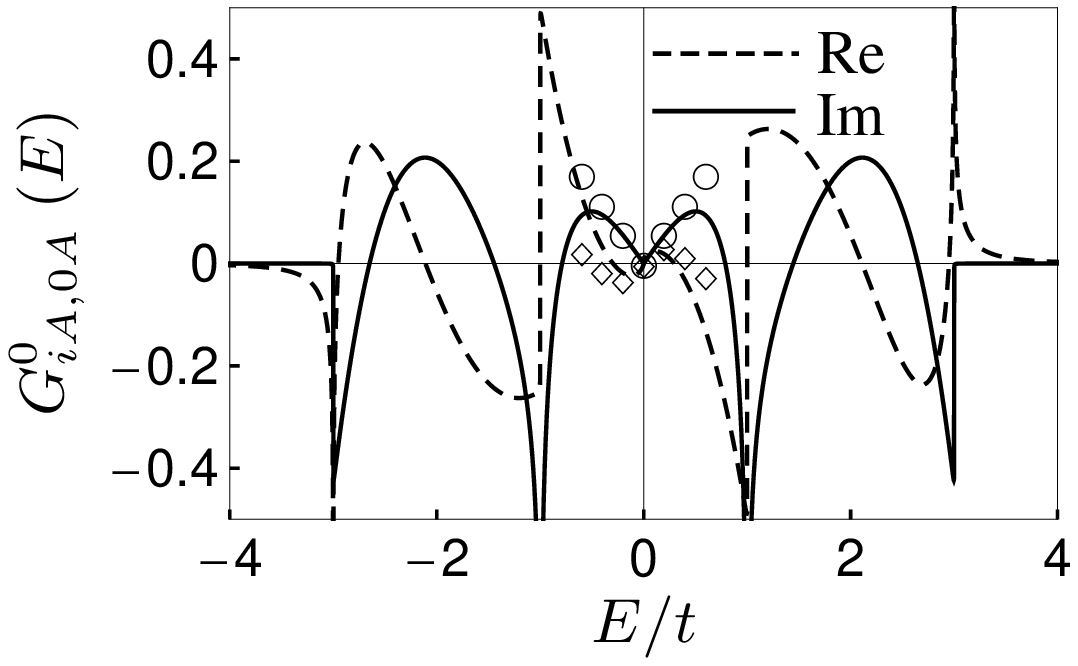}
\includegraphics[width=7.0cm]{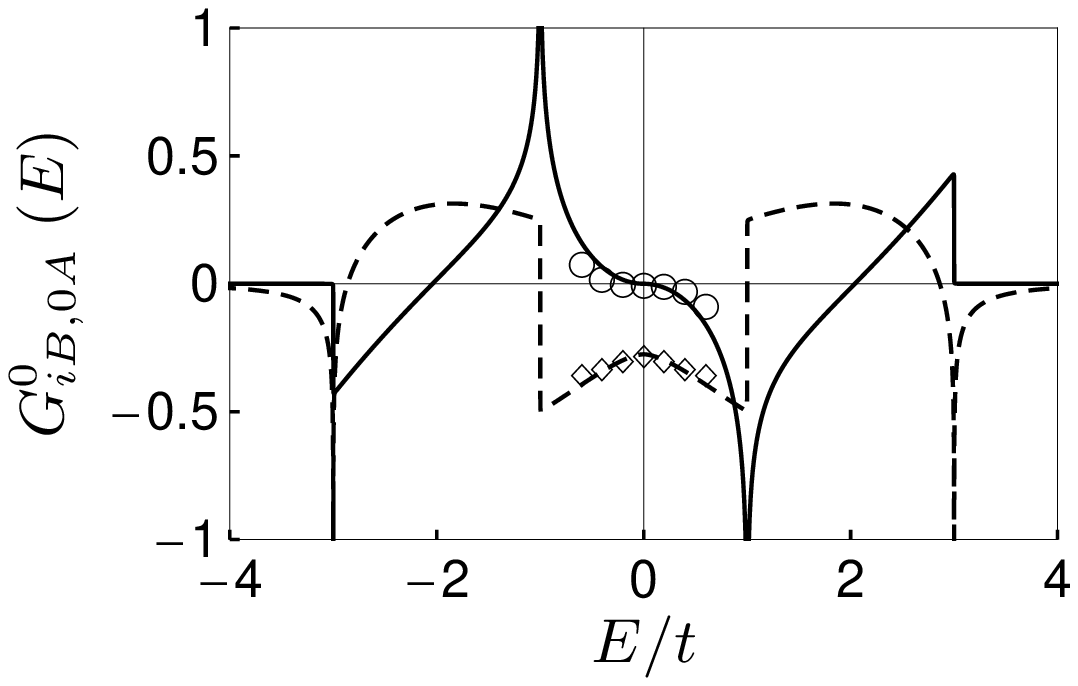}
\caption{  Illustration of the symmetry of the GF and its low energy behavior, calculated using the Horiguchi method and the tight-binding band structure. Dashed and full lines denote the real and the imaginary parts, respectively.  For the upper figure, the distance vector of the atom with respect to the impurity is given by $\vec{r} = \vec{r}_{iA} -\vec{r}_{0A} =  2 \sqrt 3 a (1, 0)$, while for the lower figure, it is $\vec{r} = \vec{r}_{iB} -\vec{r}_{0A}  = 2  a (0, 1)$, where the coordinates are indicated in Fig. \ref{spindens}. The points near $E=0$
are the low-energy  results  for the linear bands as given by Eq. \ref{GF-lowE}.
}
\label{GF-symmetry}
\end{figure}

There is no guarantee that these results, calculated for the linear bands extending to infinite energy, should agree even for low energies with the GFs for the actual bands, e.g., as obtained with the tight-binding band structure. Certain elements must exactly agree at low energies, for example, the imaginary part of  $G^0_{0 A, 0A}$, which yields the density-of-states, since at low energies, it is controlled by the Fermi velocity $v_F$ alone.
We nevertheless find that the expressions Eq. \ref{GF-lowE} do agree quite well with the tight-binding GFs, the agreement becoming better with increasing distances.
A comparison between the low-energy GFs Eq. \ref{GF-lowE} and the full GFs for a typical case is shown in Fig. \ref{GF-symmetry}, which also illustrates the symmetry of the GFs. A notable exception is the real part of the on-site GF $G^0_{0 A, 0A} (E)$, where the substitution of $r=0$ in Eq. \ref{GF-lowE} yields a divergent result. However, we find that the tight-binding GF in this case can be fitted to the expression Eq. \ref{GF-lowE} for $G^0_{0 A, 0A} (E)$,  provided we use the value $r \approx  0.6 \  a$ instead of $r = 0$.

We  note that the symmetry properties of the above GFs  Eq. \ref{GF-lowE} are consistent with the results\cite{Horiguchi} that follow from the particle-hole symmetry and valid for all energies, viz.,
$
{\rm Re} \ G^0_{i \alpha, j\beta} (E) =\mp \ {\rm Re} \ G^0_{j\beta, i \alpha} (E), \
{\rm Im} \ G^0_{i \alpha, j\beta} (E) =\pm \ {\rm Im} \ G^0_{j\beta, i \alpha} (E)$,
and
$G^0_{i \alpha, j\beta} (E) = G^0_{j\beta, i \alpha} (E)$, where the upper (lower) sign is for $\alpha = (\ne) \beta$. The symmetry properties are illustrated for specific cases in Fig. \ref{GF-symmetry}.

\begin{figure}
\centering
\includegraphics[width=7.0cm]{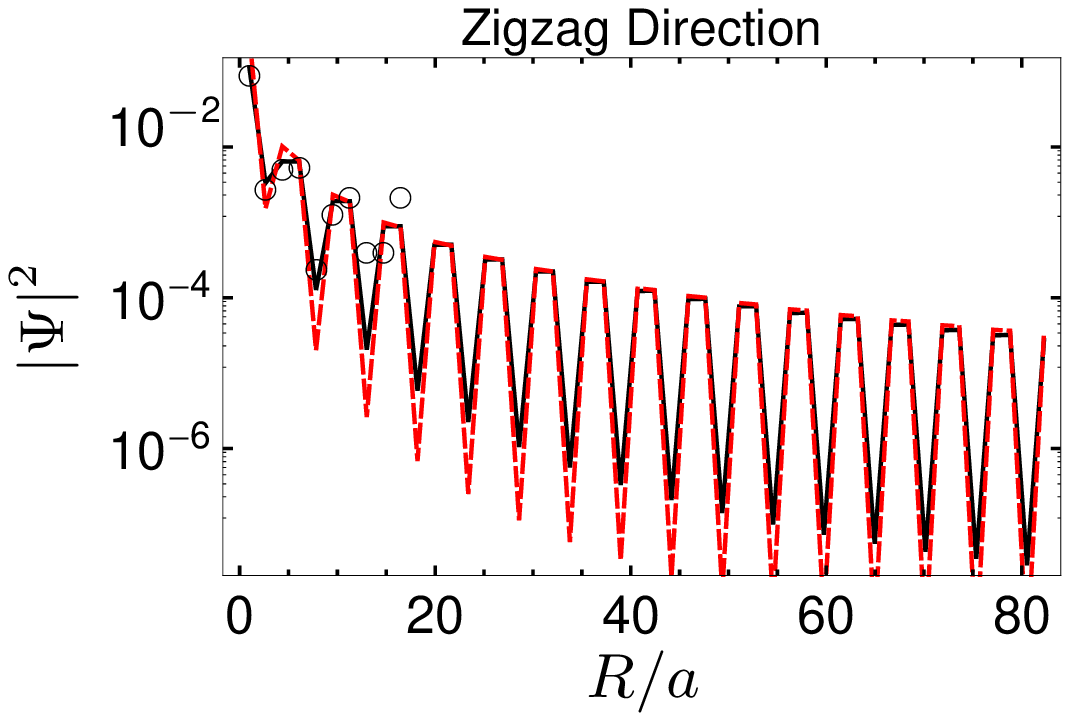}
\includegraphics[width=7.0cm]{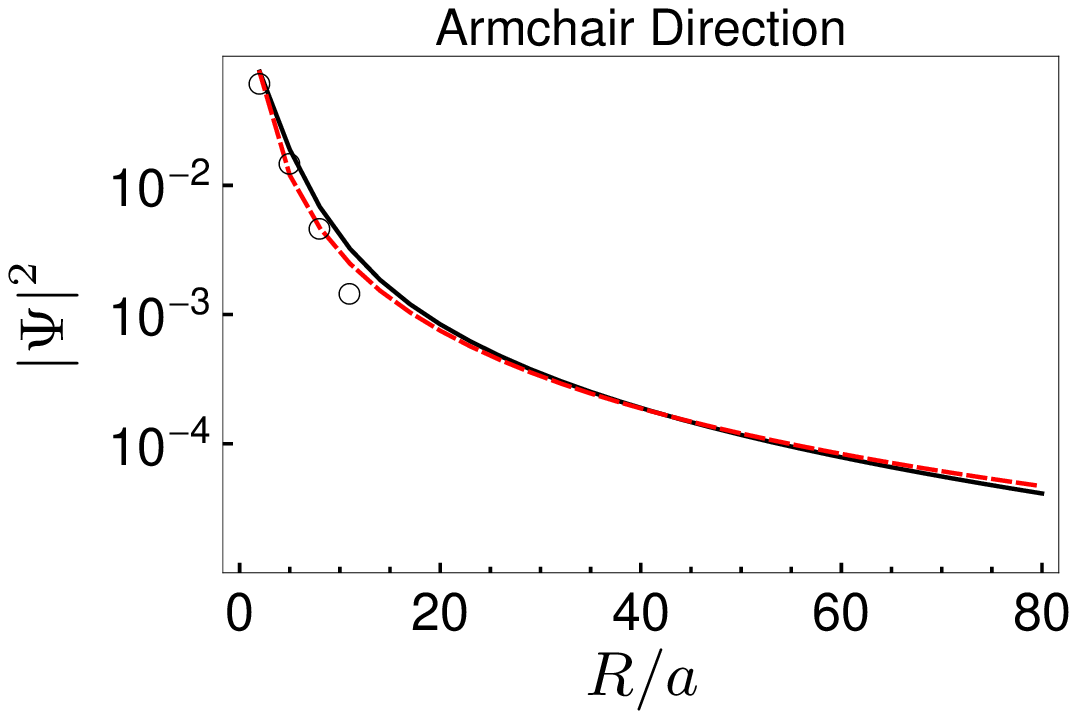}
\caption{ Square amplitude $|\Psi|^2$ of the zero-mode state on the $B$ sublattice along the zigzag   and the armchair   directions  computed from the Lippmann-Schwinger result Eq. \ref{Psi-0} using Green's Functions obtained for the (a) full TB bands  (black solid lines) and (b) linear bands  (analytical expression, Eq. \ref{Psi-zeromode})  (red dashed lines). Circles indicate the same quantity computed from the direct diagonalization of the tight-binding Hamiltonian on a finite lattice consisting of a single vacancy in a 512-atom supercell.
}
\label{psi-0}
\end{figure}

The nature of the impurity state immediately follows from the Lippmann-Schwinger expression Eq. \ref{Psi-LS}.
First of all, notice an important point from the expression for the GF Eq. \ref{GF-lowE}, viz., that all GFs vanish at $E = 0$ except for the real part of $G^0_{i B, 0A}$, which is finite and decays as $1/r$. This is precisely what leads to the property that the zero-mode state resides on the majority sublattice $B$ only and its wave function decays inversely with distance. These features are true if only the NN interactions are present on the graphene lattice. If second NN interactions are present, then there is no electron-hole symmetry and the behavior of the GFs near the resonance energy differs from Eq. \ref{GF-lowE}. The form of the GFs for the latter case is such that {\it both} sublattices contribute to the resonance state near $E=0$, an issue that is discussed in detail elsewhere.\cite{Sherafati-solidi}

 Returning to the Lippmann-Schwinger equation Eq. \ref{Psi-LS} and inserting into it the low-energy expansion for the GFs (Eq. \ref{GF-lowE}) and then taking the limit of the resonance energy $E_0 =0$,
it can be easily seen that as $E_0 \rightarrow 0$ in the limit $U_0 \rightarrow \infty$, the impurity wave function follows the behavior
  \begin{equation}
\Psi =
 \left( \begin{array}{c}
\Psi_{iA} \\
\Psi_{iB}
\end{array}\right) \
=
\left( \begin{array}{c}
1/ \ln |E_0| \\
c_i |E_0| ^{-1}
\end{array}\right)
\rightarrow
\left( \begin{array}{c}
0 \\
c_i
\end{array}\right).
\label{Psi}
\end{equation}

This is an important result, which states that in the NN approximation, only the $B$ sublattice component survives for the zero-mode state, it being the stronger infinity.
 The surviving component  is found to be simply proportional to the real part of the inter-sublattice GF,
\begin{equation}
\Psi_{iB} \propto   {\rm Re} \ G^0_{i B, 0A} (E_0 \rightarrow 0),
\label{Psi-0}
\end{equation}
since its imaginary part vanishes. Using Eq. \ref{GF-lowE}  and evaluating $\rm Im \  \alpha$ from
Eq. \ref{alpha}, we finally get the desired result
\begin{equation}
\Psi_{B} (r)  = \frac{N} {r} \sin[(\vec{K}-\vec{K'})\cdot \vec{r}/2-\theta_r]\cos[(\vec{K}+\vec{K'})\cdot \vec{r}/2-\pi/3],
\label{Psi-zeromode}
\end{equation}
where we have suppressed the cell index $i$, $N$ is a constant, $r$ is again the actual distance vector of the $B$ site with respect to the impurity position, and the two Dirac points in the Brillouin zone may be taken as:
$K = 2\pi a^{-1} 3^{-3/2} (-1, \sqrt{3})$ and $K' = 2\pi a^{-1} 3^{-3/2} ( 1, \sqrt{3})$.

Eq. \ref{Psi-zeromode} is the central result of this Subsection that describes the $1/r$ decay of the vacancy-induced V$\pi$ state along with the phase factors. The long-range nature $1 / r$ of  the wave function (\ref{Psi-zeromode}) is well-known,\cite{Pereira2} but the oscillatory factor due to the interference effect of the two Dirac points is new. The same kind of interference is also present in the oscillations of the
RKKY interactions.\cite{Saremi, Sherafati}
The wave function is not square integrable because we used the linear band structure, but it will be if we take the full band structure into account.  Eq. \ref{Psi-zeromode} nevertheless describes the gross features of the zero-mode state. The wave function changes sign along different directions, e. g., it changes sign along the zigzag direction but not along the armchair direction. The kinetic energy gained by the delocalization of the wave function is exactly cancelled by anti-bonding components present in the wave function, so that its energy still equals the on-site energy in spite of the delocalization. 	The calculated wave function for the zero-mode state is shown in Fig. \ref{psi-0}. We note that a recent study has shown that the $1/r$ decay of the vacancy state remains unchanged even when a repulsive Coulomb interaction is included in the tight-binding Hamiltonian.\cite{Haas}

\section{Summary}
In summary, we have studied the electronic structure of graphene with a single substitutional vacancy from density-functional calculations using the all-electron LAPW method and interpreted the results with the help of the tight-binding model and the impurity Green's Function approach. We find that the vacancy induces four localized states, viz., three V$\sigma$ dangling bond states on the carbon triangle and one V$\pi$ resonance state. The dangling bond states cause a Jahn-Teller distortion, which we found to be a planar distortion of the carbon triangle. Hund's coupling between these electrons would then produce the $S=1$ state at the vacancy center as indicated in the summary figure Fig. \ref{sketchdos}. The magnetic moment has two components: (i) The component $ 2 \mu_B$ coming from the localized vacancy states V$\sigma$ and V$\pi$ and (ii) An opposite component of several tenths of $\mu_B$ coming from the spin-polarization of the continuum $\pi$ band states in the vicinity of the vacancy. The second part is  not well described in  the supercell band calculations due to the slow $1/r$ decay of the ``quasi-localized" V$\pi$ wave function.
This long-range decay also means that in an experimental sample it is only for the extremely low vacancy concentration that the truly isolated vacancy limit is reached and as a result the magnetic moment is expected to be dependent on the vacancy concentration.

In addition to the density-functional calculations, we also studied the formation of the V$\pi$ state in detail from the impurity Green's function approach for the isolated vacancy, which provided important insight in the interpretation of the results of the band calculations and the formation of the zero-mode states in the $\pi$ bands.
This zero mode state is a slowly-decaying localized state that lives mostly on the majority sublattice. It spreads into the minority sublattice (the one containing the vacancy) and becomes a resonance state due to the second and the higher-neighbor interactions as well as the finite strength of the vacancy potential. The Green's function approach provided a sinusoidal phase factor associated with the V$\pi$ wave function described by Eq. \ref{Psi-zeromode}.
In addition to the understanding of the vacancy electronic structure, our work provides important insight necessary for the understanding of impurities in general such as iron and cobalt dopants and other complex defects.

\ack
This work was supported by the U. S. Department of Energy through Grant No.
DOE-FG02-00ER45818.


* Permanent Address: Department of Physics, Indian Institute of Technology Madras, Chennai 600036, India

** Permanent Address: Institute of Nuclear Sciences, Vin\v{c}a, University of Belgrade, P. O. Box 522, 11001 Belgrade, Serbia

\end{document}